# Nanoscale magnetophotonics

Nicolò Maccaferri[1*], Irina Zubritskaya[2], Ilya Razdolski[3], Ioan-Augustin Chioar[4], Vladimir Belotelov[5], Vassilios Kapaklis[4], Peter M. Oppeneer[4] and Alexandre Dmitriev[6*]

[1]Department of Physics and Materials Science, University of Luxembourg, 162a avenue de la Faïencerie, L-1511, Luxembourg, Luxembourg

[2]Geballe Laboratory for Advanced Materials, Stanford University, 476 Lomita Mall, Stanford, California 94305-4045, USA

[3]Fritz Haber Institute of the Max Planck Society, 14195 Berlin, Germany

[4]Department of Physics and Astronomy, Uppsala University, P. O. Box 516, S-75120 Uppsala, Sweden

[5]Lomonosov Moscow State University, Moscow 119991, Russia

[6]Department of Physics, University of Gothenburg, S-412 96 Gothenburg, Sweden

[*]nicolo.maccaferri@uni.lu

[*]alexd@physics.gu.se

This Perspective surveys the state-of-the-art and future prospects of science and technology employing the nanoconfined light (nanophotonics and nanoplasmonics) in combination with magnetism. We denote this field broadly as nanoscale magnetophotonics. We include a general introduction to the field and describe the emerging magneto-optical effects in magnetoplasmonic and magnetophotonic nanostructures supporting localized and propagating plasmons. Special attention is given to magnetoplasmonic crystals with transverse magnetization and the associated nanophotonic non-reciprocal effects, and to magneto-optical effects in periodic arrays of nanostructures. We give also an overview of the applications of these systems in biological and chemical sensing, as well as in light polarization and phase control. We further review the area of nonlinear magnetophotonics, the semiconductor spin-plasmonics, and the general principles and applications of opto-magnetism and nano-optical ultrafast control of magnetism and spintronics.

# I. Introduction

During the past two decades our ability to control materials at the nanoscale allowed a more aware study of nanoscale light-matter interactions, leading to the advent of nanophotonics and nano-optics. One particularly efficient way of confining light into subwavelength volumes is by using the collective electromagnetic-induced electronic excitations known as plasmons. Unlike conventional optics, plasmonics enables the unrivalled concentration and enhancement of electromagnetic radiation well beyond the diffraction limit of light [1-4]. Besides its fundamental scientific importance, manipulation of light at the nanoscale is of great interest due to its potential exploitation towards real-life applications such as energy harvesting and photovoltaics, wave-guiding and lasing, optoelectronics, biochemistry and medicine.

To achieve new functionalities, the combination of plasmonics with other material properties has become increasingly appealing. In particular, magnetoplasmonics and magnetophotonics are emerging areas that aim at combining magnetism, plasmonics and photonics [5-11] to find new ways of controlling the properties of plasmons using magnetic fields or vice-versa, to control magnetic properties with light. Nanoscale magnetophotonics entails the fundamental studies of photon–electronic spin interactions in nanostructured materials [12]; the enhancement of magneto-optical (MO) activity in materials [13-15], including dielectrics [16], 2D materials [17], nanoparticle-decorated graphene [18] and graphene-based metasurfaces and their topological transformations [19]; the active control of plasmons with weak magnetic fields [20]; topological photonics and gyromagnetic photonic crystals [21]; magnetoplasmonics-based bio- and chemical sensing [22] and magnetophotonic and magnetoplasmonic crystals (MPCs) as modulators of light transmission, reflection and polarization [23-25].

Since the early 1970s, the investigation of the interaction between magnetism and plasmons has been a topic of high interest. In 1972 Chiu and Quinn showed that an external static magnetic field could control the properties of surface plasmon polaritons (SPPs) such as their propagation or localization [26] [Fig. 1(a)]. The exponential growth of fabrication techniques in semiconductor technology during the last two decades boosted engineering of photonic band gap materials and plasmonic systems operating at optical frequencies. MO properties of photonic crystals and their potential use in integrated optics were thoroughly investigated in 2005 by Belotelov and Zvezdin [27]. Shortly after, extraordinary transmission and plasmon-enhanced giant Faraday and Kerr effects were demonstrated in noble metal-dielectric plasmonic system made of Au films with either a sub-wavelength hole [28] or slit [29,30] array on top of a magnetic Bi:YIG layer. In parallel, the theoretical study by Yu *et al.* predicted that a waveguide formed at the interface between a photonic crystal and a metal under a static magnetic field possesses unique dispersion relations resulting in modes propagating in only one allowed direction [31]. In 2007 Gonzalez-Diaz et al. demonstrated that the coupling of an external magnetic field to the surface propagating plasmon wave vector can be greatly enhanced in noble metal/ ferromagnetic/ noble metal trilayers, which allows magnetic control of surface plasmon analogously to semiconductors [32]. At the end of 1990s, Martín-Becerra et al. showed that magnetic modulation of SPP wave vector could be significantly improved by depositing a dielectric overlayer in such geometries [33] and, later on, that both the real and imaginary parts of SPP wave vector are affected by the magnetic field in noble/ ferromagnetic/ noble metal films resulting in spectrally dependent modulation [34]. Unfortunately, for noble metal-based plasmonic structures, the magnetic field required to achieve proper control of surface plasmon properties is too high for application purposes. With nanoengineering of complex systems combining ferromagnetic materials and noble metals, which exhibit simultaneously magnetic and plasmonic properties, it became possible to control the plasmon wave vector with a weak (100 mT regime) external magnetic field [35,36], generate ultrashort SPP pulses [37] and produce SPP-induced magnetization in nickel with effective magnetic field of 100 Oe by femtosecond laser pulse [38]. Hybrid magnetoplasmonic systems combining noble metal and iron garnets that are typically highly transparent compared to ferromagnetic metals provide magnetic modulation of light transmittance. Enhanced MO effects and strong magnetic modulation of light intensity were found in metallic nanostructures integrated with iron garnet film [39-43]. Furthermore, plasmon mediated MO transparency was observed in magnetophotonic crystal formed by gold grating stacked on top of bismuth-substituted rare-earth iron garnet deposited on top of gadolinium gallium garnet [25]. In similar architectures, a shift of plasmon polariton resonance was manipulated by femtosecond laser pulses [44]. Finally, in systems that combine plasmonic crystals and magnetic semiconductors the MO effects could be dramatically enhanced in both transmission and reflection [45].

In parallel to the studies on propagating plasmons, the current rapid advances in nanofabrication enable the broadening of our understanding of optics at the nanoscale with nanostructures supporting also localized surface plasmon resonances (LSPRs) [Fig. 1(b)]. Also here hybrid structures were proposed to combine all-in-one the advantages of noble-metals and magnetic materials [46,47]. Also, pure ferromagnetic nanostructures were demonstrated to support LSPRs [48] and SPPs [49,50] and at the same time exhibit sizeable magnetic effects under low magnetic fields, leading to a large tunability of the MO response [51]. The strong coupling between SPP and the MO activity leading to a significant enhancement and tunability of the Kerr effect as a result of lattice design were observed also in pure ferromagnetic (Fe, Co, Ni) 2D hexagonal lattices [52-54].

MO effects can be classified into i) the Faraday effect giving rise to rotation and ellipticity of an incoming linearly polarized light transmitted through a magnetized medium, and ii) the MO Kerr effect (MOKE), which induces similar effects but in reflection configuration (see Ref. [55] for a theoretical description of MO effects). Different configurations of the MOKE can be defined depending on the relative orientation between the magnetization vector **M**, the plane of incidence of the incoming light and the reflective surface [Fig. 1(c)]. In longitudinal MOKE (L-MOKE) configuration, **M** lies both in the plane of the sample and in the plane of incidence of the incoming light [Fig. 1(c)]. In polar MOKE (P-MOKE) geometry, **M** is oriented perpendicularly to the reflective surface and parallel to the plane of light incidence [Fig. 1(c)]. Finally, in the transverse MOKE (T-MOKE) configuration, **M** is lying on the sample surface and oriented perpendicularly to the plane of incidence of the incoming light [Fig. 1(c)].

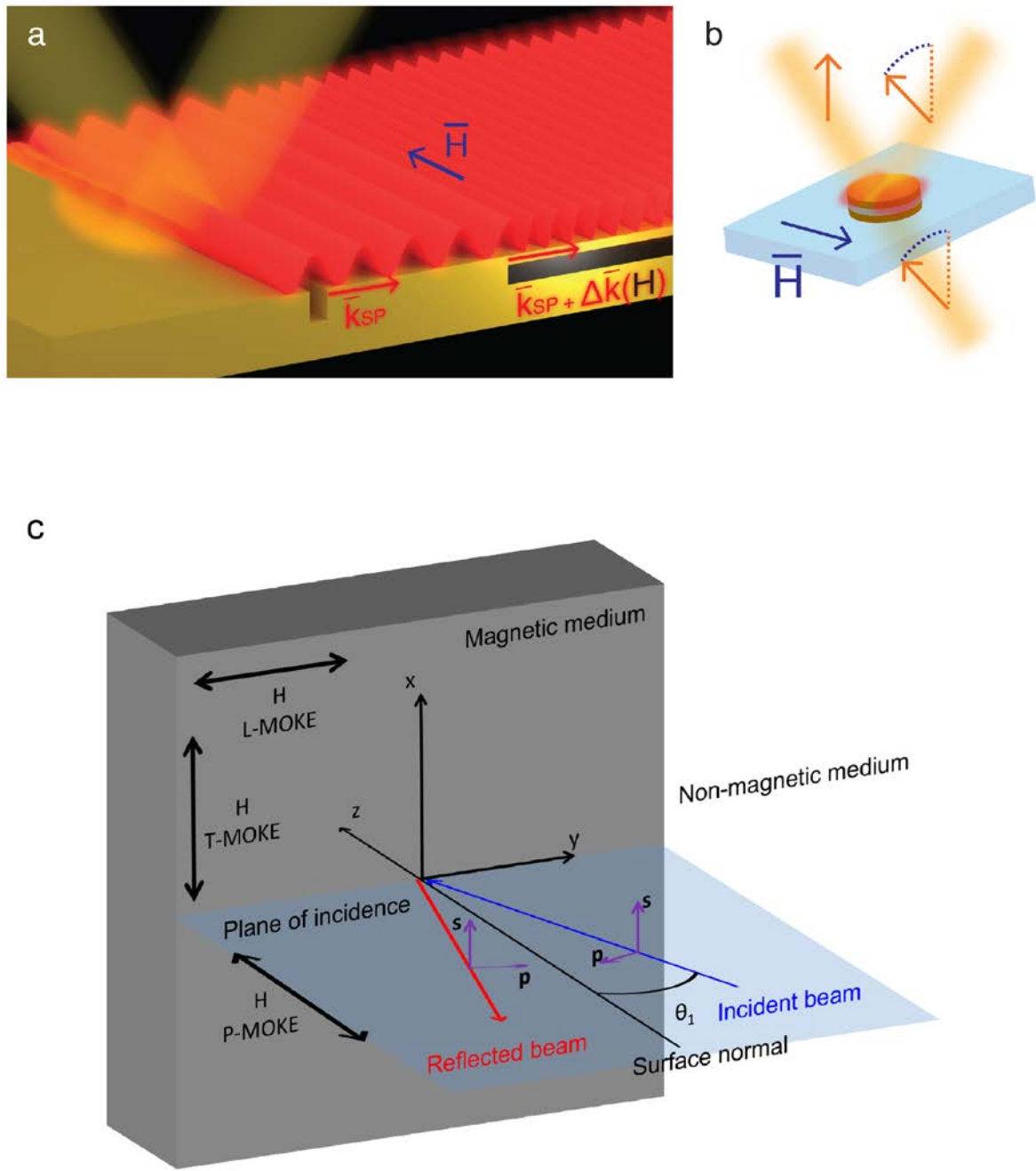

**Fig. 1.** (a) Modulation of a SPP, launched by the groove, by an external magnetic field **H** when a ferromagnetic layer (black) is inserted into the noble metal film (yellow); (b) sketch of Kerr (reflection) and Faraday (transmission) effects in multilayer noble metal/ferromagnet nanoantenna supporting a LSPR. The polarizations of the incoming and outcoming light are shown, including the magnetic-field-induced and plasmon-enhanced polarization rotation (the applied in-plane external magnetic field is marked). The excited LSPR in the nanoantenna is highlighted by the dipolar near-field pattern; (c) MOKE configurations and coordinate system used in their description.

This Perspective covers a plethora of intriguing effects and phenomana associated with light-matter interactions in nanoscale geometries in the presence of magnetic field. Section I contains a brief introduction to the field of magnetophotonics and highlighs important discoveries in geometries supporting propagating and localized plasmons. For smooth navigation through this Perspective, the reader can always refer to the classification of the MOKE configurations that is given in Fig. 1(c) of Sec. I. Section II of this Perspective mainly covers MO effects in magnetoplasmonic and magnetophotonic nanostructures in different configurations. We start with Part (a) of Sec. II, that presents the overview of fundamental works that theoretically explored the origin of MO by analytical models and explained the role of spin-orbit (SO) coupling in the MO activity in nanostructures supporting localized plasmons. We then proceed to Sec. II B where we explain the origin and the resonant enhancement of MO effects in MPCs and derive the dispersion relations. In both Secs. II B and II B we discuss the fundamental limitations and the main strategies used to maximize the MO enhancement in magnetoplasmonic nanostructures and MPCs. In Sec. II C we discuss transversely magnetized MPCs and plasmonic nonreciprocity, specifically focusing on the variety of materials and geometries that provide strong light modulation by the transversely applied magnetic field. Section II D delves into MO effects in longitudinal magnetization and introduces the longitudinal magnetophotonic intensity effect (LMPIE). Finally, we devote Sec. II E to MO effects in dot- and antidot periodic arrays and consider special light illumination conditions associated with Wood's anomalies and second harmonic generation. In Sec. III we give an overview of applications of nanoscale magnetoplasmonics and magnetophotonics in biological and chemical sensing and light's polarization and phase control. Section IV is entirely focused on nonlinear-optical processes attainable in the vicinity of SP resonances in the presense magnetic fields. We continue with Sec. V that introduces the emerging field of magnetically induced spin-polarization in semiconductors. Section VI of this Perspective is devoted to ultrafast magnetism and fundamental understanding of the relationship of spin orbital momentum and orbital angular momentum of light and nanoscale magnetism giving a special attention to the inverse Faraday effect and helicity-dependent all-optical magnetization switching. We conclude by giving our outlook on the field and by summarizing the recent advances that pave the way to practical magnetophotonic devices.

## II. Magneto-optical effects in magnetoplasmonic and magnetophotonic nanostructures

### a. Localized plasmons in magnetoplasmonic nanostructures

Magnetoplasmonic nanostructures and nanostructured magnetophotonic crystals support surface plasmon resonances (localized and/or propagating). Therefore, they exhibit strongly enhanced MO activity at low magnetic fields. Regarding the systems supporting LSPRs, Sepulveda et al. first explained intuitively this phenomenon in 2010 [13]. They showed that in pure gold nanodisks the large MO response comes from an increase of the magnetic Lorentz force induced by the large collective movement of the conduction electrons when a LSPR is excited in the presence of a static magnetic field [see Figs. 2(a) and 2(b)].

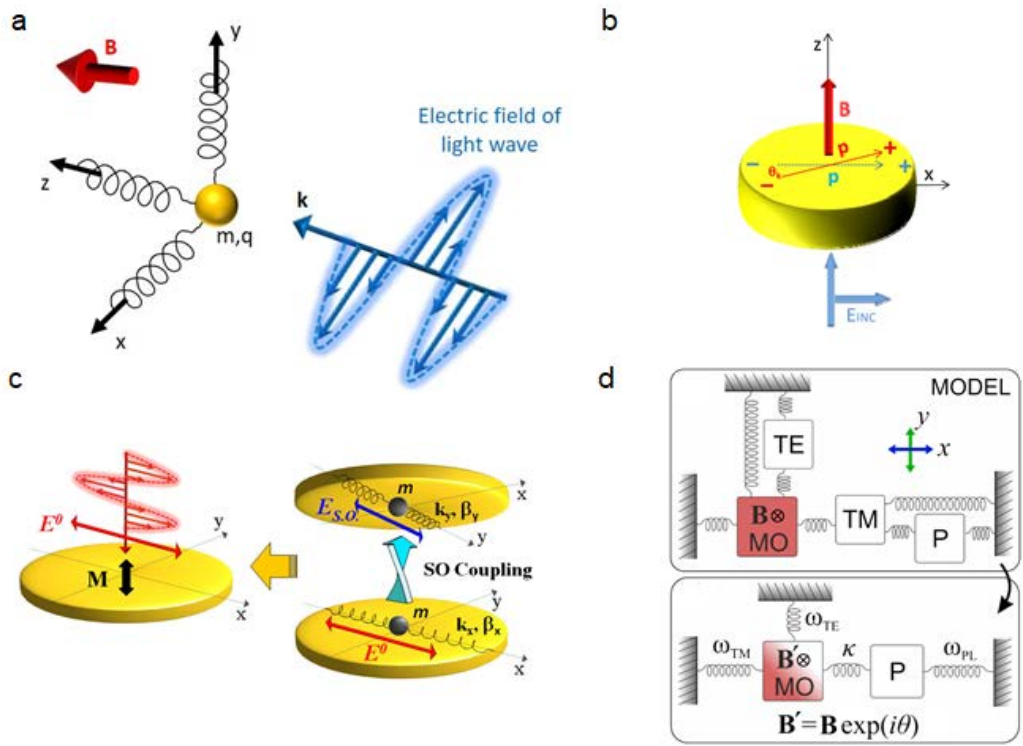

**Fig. 2.** (a) Schematic drawing of the mechanical oscillator model for magneto-optic solids. It corresponds to the standard Lorentz oscillator model for dielectrics, but with the addition of a static magnetic field, which exerts a Lorentz force on the bound electrons (adapted from Ref. [9]). (b) Schematic of the MO effect induced by the Lorentz force in a metal nanoparticle [13]. (c) A ferromagnetic disk modeled with two orthogonal damped harmonic oscillators coupled by the SO interaction; m represents the mass of the conduction electrons; the spring constants $k_x$ and $k_y$ originate from the electromagnetic restoring forces due to the displacements of the conduction electrons; $\beta_x$ and $\beta_y$ are the damping constants [56]. (d) Top-panel: mechanical analog that represents the coupling of the relevant optical excitations; bottom-panel: simplified oscillator model providing analytical solutions [57]. Copyright 2010 American Physical Society. Copyright 2013 American Physical Society. Copyright 2016 American Physical Society.

Few years later, Maccaferri et al. [56] provided a semi-classical explanation by exploring the influence of the phase of localized plasmon resonances on the MO activity in nickel nanodisks. They demonstrated that these systems can be described as two orthogonal damped oscillators coupled by the SO interaction, proving that only the SO-induced transverse plasmon plays an active role on the MO properties by controlling the relative amplitude and phase lag between the two oscillators [Fig. 2(c)]. Furthermore, a full analytical theoretical description for typical sample geometries was introduced recently by Floess et al. [57], who developed a Lorentz nonreciprocal coupled oscillator model [Fig. 2(d)] yielding analytical expressions for the resonantly enhanced MO response. All these models can be transferred to other complex and hybrid nano-optical systems and can significantly facilitate device design. However, the magnetic field-induced modulation of light polarization achieved in magnetophotonic crystals so far is only in the order of a fraction of degree, which is insufficient for any practical purposes. When using conventional ferromagnets, the main obstacles are the exiguity of MO activity arising from the SO coupling and the rather inefficient excitation and/or propagation of plasmon modes, due to their high dissipative losses. One of the key challenges is indeed to increase the strength of SO-coupling without increasing the plasmon damping. The main strategies currently pursued with conventional ferromagnetic materials, namely without increasing the intrinsic SO-coupling, are (i) periodic arrangements of magnetoplasmonic nanoantennas [58,59];

(ii) 3D ferromagnets [60] and composite ferromagnetic/noble metal [61] and ferromagnetic/dielectric/noble metal nanostructures [62], and (iii) heterogeneous units comprising multiple nanoantennas placed in proximity to enable their near-field interaction [63-67]. Initial investigations have shown that the enhancement of polarization rotation by one order of magnitude can indeed be achieved following these strategies. Finally, it is worth noticing that exploting high-index all-dielectric nanostructures one can reduce the high losses, which are inherent in magnetic materials [16]. The use of these materials can lead to peculiar novel phenomena where magnetic dipoles are responsible for the MO activity, thus opening interesting perspectives in the engineering of novel nanoscale MO effects.

### b. Magnetoplasmonic crystals

Periodically nanostructured metal-dielectric systems allow excitation of propagating plasmonic modes by incident light. On the other hand, their periodicity is of the order of wavelength of SPPs propagating at the metal-dielectric interface and at some frequency range constructive interference takes place and band gaps appear. Therefore, such kind of structures can be referred to the plasmonic crystals in analogy to photonic crystals. If some magnetic substances are involved, then such periodic structure is called *magnetoplasmonic crystal* (MPC).

There are several designs of MPCs including one dimensional (1D) [Figs. 3(a) and 3(b)], two-dimensional (2D) [Figs. 3(c)–3(f)] crystals.

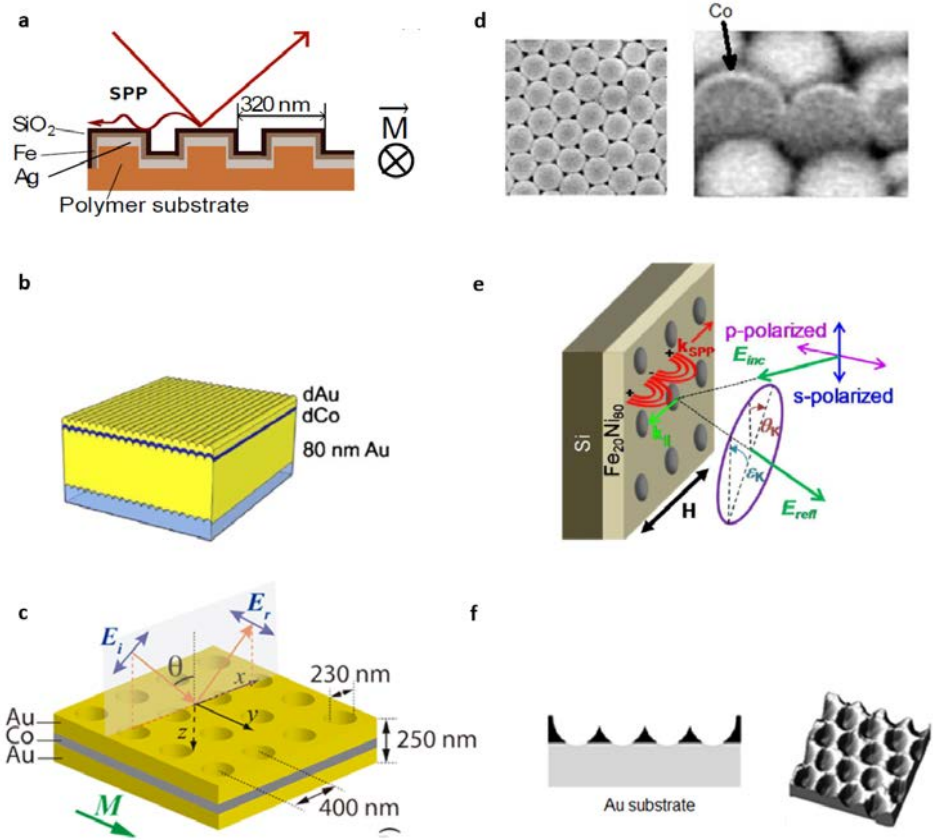

**Fig. 3.** Different types of MPCs. (a) 1D trilayer SiO2/Fe/Ag MPC fabricated on a blue-ray disc. nickel grating. Reprinted with permission fro Ref. [71]. Copyright 2016 Elsevier. (b) Trilayer Au/Co/Au grating on a polycarbonate grating. Reprinted with permission form Ref. [72]. Copyright 2010 The Optical Society. (c) 2D MPC in trilayer of Au/Co/Au. Reprinted with permission from Ref. [79]. Copyright 2016 American Chemical Society. (d) 2D nanocorrugated magnetic film of cobalt on the top of PMMA colloidal crystal: (left) SEM image and (right) microphotography of the particles cross-section made by focused Ga ion beam, the Co coverage is visible as a bright layer. Reprinted with permission from Ref. [77]. Copyright 2010 The Optical Society. (e) 2D MPC of permalloy on Si substrate. Reprinted with permission from Ref. [49]. Copyright 2015 American Chemical Society. (f) 2D plasmonic crystal from self-assembled polymeric monolayers replicated on nickel on a gold substrate: (left) schematics and (right) AFM image [78] Copyright 2011 American Institute of Physics.

Generally, excitation of plasmonic resonance provides enhancement of MO effects. A noble metal plasmonic crystal without any magnetic media can be also made MPC if a high external magnetic field is applied. If the magnetic field is in-plane and transverse with respect to SPP propagation then it provides some enhancement of the T-MOKE [68]. In this case the MO properties are due to Lorentz force acting on free electrons in a magnetic field. A resonant increase of the T-MOKE was reported for one-dimensional Co, Fe and Ni gratings [69,70]. Pronounced resonance of T-MOKE in a sample of 1D trilayer SiO2/Fe/Ag MPC fabricated on a commercial blue-ray disc also allowed to consider a refractive index sensor on its basis [71] [Fig. 3(a)]. Though propagation length of SPPs in ferromagnetic metals is rather small and does not exceed several microns, it still counts several periods of the structure and the periodicity plays an important role in the SPPs excitation and their interplay with MO effects. Several times increase of the T-MOKE at the plasmonic resonances of the grating with respect to the smooth ferromagnets was reported. The concepts of hybrid MPC based on noble-metal/ferromagnetic-metal multilayers as well as nanocorrugated 2D films were also comprehensively studied in [72-76] [Figs. 3(b) and (c)] and in [77,78] [Figs. 3(d) and 3(f)], respectively. Since the overall optical losses for such systems are lower than for pure ferromagnetic metals the effect of resonant increase of the T-MOKE due to propagating SPPs in these structures is more pronounced. It also allows to consider these structures as highly sensitive plasmonic biosensors [71,79]. Concept of MPC works not only with transverse magnetization. Recently, Maccaferri et al. investigated longitudinally magnetized MPC and observed increase of the L-MOKE at the SPP resonances [49] [Fig. 3(e)].

The MPC structures can also be referred as magnetophotonic metasurfaces, though the term of metasurface is more general and also includes all-dielectric and semiconductor materials consisting of substrates covered with cylinders and spheres sustaining Mie resonances [80,81]. An example of magnetoplasmonic metasurface is represented by two-dimensional arrays of Si nanodiscs covered by a thin Ni film [82]. Optical resonances in such samples lead to enhanced MO response like Faraday rotation of 0.8 deg. which is reasonably large taking in mind that the magnetic part is only 5 nm thick.

The main disadvantage of most of the aforementioned approaches is that the optical losses associated with the presence of a ferromagnetic metal are still relatively high. This fact limits exploiting fully the potential gain of the combined concepts of nanostructuring and plasmonics in magneto-optics. If the ferromagnetic metals were avoided as in cases of pure semiconductors or noble metal systems, huge external magnetic fields exceeding several Tesla would be necessary to make the T-MOKE at least comparable with the effect in ferromagnets. That is why it seems that the plasmonic crystals containing low-loss ferromagnetic dielectrics and noble metals can provide even better results [9,14,83]. The most pronounced enhancement of the MO effects takes place for high-quality resonances that are achieved if the ferromagnetic metal is substituted by a low absorptive noble one and the dielectric layer is magnetized. Probably, the best candidates for magnetic dielectric are bismuth rare-earth iron garnet films of composition $Bi_xR_{3-x}Fe_5O_{12}$, where R is a rare-earth element [84]. Therefore, we will consider main properties of MPCs taking this kind of structures as exemples and study their properties in detail.

Let us consider an MPC consisting of smooth magnetic dielectric on a substrate and noble metal film periodically perforated with subwavelength array(s) of slits and holes. In such structure SPPs can propagate either along the upper interface, the air/metal interface, or along the bottom interface between the metal and magnetic dielectric. Though the metal film is not continuous, the SPP can still propagate along the structure if the air gap size is notably smaller than the SPP wavelength and air takes relatively small part of the MPC crystal lattice. During SPP propagation some part of its energy continuously leaks in the far-field due to the SPP scattering on the metal grating. This mechanism also contributes to the SPP energy decrease in MPCs together with conventional energy dissipation in lossy metal and dielectric layers.

On the other hand, metal perforation provides a very efficient way to excite SPPs by using light. The metal grating provides diffracted light with different in-plane wavevector components. If some of them coincide with the SPP wavevector then the light will be coupled to SPPs. In this case the momentum conservation law is written as

$$k_0\sqrt{\varepsilon_3}\sin\theta \mathbf{e}^{(in)} = \beta \mathbf{e}_{SPP} + u_1 \mathbf{G}_x + u_2 \mathbf{G}_y, \qquad (1)$$

where $k_0$ is the wavevector of light in vacuum, $\beta$ is the SPP wavenumber along the metal-dielectric interface, $\varepsilon_3$ is the dielectric constant of the medium above the metal/dielectric structure, $\theta$ is the angle of incidence, $\mathbf{G}_x$ and $\mathbf{G}_y$ are two reciprocal lattice vectors, $|\mathbf{G}_x| = 2\pi/d_x$, $|\mathbf{G}_y| = 2\pi/d_y$; $d_x$ and $d_y$ are the periods of the grating along the *x*- and *y*-directions; $\mathbf{e}^{(in)}$, $\mathbf{e}_{SPP}$ are two in-plane unit vectors along the plane of light incidence and along the SPP propagation direction, respectively, and $u_1$ and $u_2$ are integers. In the grating configuration, SPPs can be excited on both the metallic interfaces.

Strictly speaking, the absolute value of the wavevector $\beta$ of the grating SPP in Eq. (1) deviates from the one for the smooth metal-dielectric interface determined by $= k_0 \sqrt{\frac{\varepsilon_1 \varepsilon_2}{\varepsilon_1 + \varepsilon_2}}$, where $\varepsilon_1$ and $\varepsilon_2$ are dielectric permittivites for the metal and dielectric, respectively. In the case of a metal grating with narrow slits/holes on a smooth dielectric this deviation is usually rather small and formulas for smooth interfaces are well applicable. However, the periodicity of the slits/holes does not allow describing SPP dispersion fully by the effective medium approach. This becomes mostly pronounced at $k = u(\pi/d)$ with an integer *u*, where the dispersion curve splits into two branches: low and high frequency and a band gap appears. This phenomenon is a general feature of any periodic structure with a period comparable with the wavelength of the wave propagating through it. Such periodic structures dealing with photons are called photonic crystals. That is why periodic metal-dielectric structures considered here can be referred to as plasmonic crystals. Plasmonic crystals allow tailoring dispersion of SPP in a desired way and concentration electromagnetic energy in a small volume near the metal/dielectric interface. The latter was shown recently to have a great potential for ultrafast nanophotonics since it allows switching permittivity of gold by a short laser pulse at a time scale of several hundreds of femtoseconds [44].

It should be noted that most of the results on magnetoplasmonics-assisted light control were obtained at visible and near IR spectral range. However, modern telecommunication technologies rely on 1.55 µm light. At this wavelength plasmonic properties of noble metals still remain relevant while one should be careful about the choice of a magnetic dielectric. In particular, bismuth substituted rare-earth iron garnets have rather low MO activity at 1.55 µm. In this case, the use of cerium substituted iron-garnets seems to be more preferable [85].

### c. Transversely magnetized magnetoplasmoni crystals and plasmonic nonreciprocity

An MPC can be magnetized in different directions by external magnetic field. It follows from Maxwell's equations that the T-MOKE does not change the polarization state of the SPP but only its wavenumber β. In this configuration the vector product of the magnetization **M** and the vector **N** normal to the interface is nonzero near the surface of the magnetized medium (e.g. a thin film). The magnetic field breaks the time-reversal symmetry, while the presence of an interface and a normal vector associated with it breaks the spatial inversion. Interestingly, the space-time symmetry breaking is characteristic of media with a toroidal moment τ whose transformation properties are identical to those for **M** × **N** [86]. Thus, the propagation of SPP is similar to the propagation of a wave in a medium with a toroidal moment along its direction. In electrodynamics, the presence of a toroidal moment is known to give rise to optical nonreciprocity. In the case under consideration, the latter is manifested in a difference between the wave vectors of the electromagnetic wave as it propagates in the direction along the vector τ and in the opposite direction [87]:

$$\kappa = k_0\sqrt{\varepsilon}(1 + \frac{(\tau \cdot \mathbf{k}_0)}{k_0\sqrt{\varepsilon}}) \qquad (2)$$

Similar optical nonreciprocity takes place for a SPP in the case of a transversally magnetized medium

$$\kappa = \kappa_0(1 + \alpha \mathrm{g}) \qquad (3)$$

where $\kappa_0 = k_0(\varepsilon_1\varepsilon_2/(\varepsilon_1+\varepsilon_2))^{1/2}$ and $\alpha = (-\varepsilon_1\varepsilon_2)^{-1/2}(1-\varepsilon_2^2/\varepsilon_1^2)^{-1}$; $\varepsilon_1$ and $\varepsilon_2$ are the dielectric constants of metal and dielectric, respectively, and gyration *g* is a parameter linear in the magnetization that is responsible for the MO properties of the material (in terms of the dielectric tensor, $g=i\varepsilon_{zx}=-i\varepsilon_{xz}$ if the magnetization is directed along y-axis). It follows from Eq. (2) that, in the first approximation, the wavenumber of the surface wave depends linearly on the film gyration *g*, which confirms the nonreciprocity effect. Equation (3) agrees with Eq. (2) if it is considered that, according to what has been said so far, $\tau \sim \mathbf{M} \times \mathbf{N}$ and, hence, $g_y \sim \tau_x$.

Excitation of the SPP influences optical transmission and reflection spectra of the MPC providing asymmetric shape (so-called Fano-shaped resonances). The magnetization-induced changes in SPP dispersion shifts the Fano resonances and the MO intensity shift appears.

In an MPC where the metallic layer is perforated by an array of parallel slits and the dielectric layer is magnetized along the slits [Fig. 4(a)], the plasmonic resonances in reflection and transmission are shifted in frequency depending on the magnitude and direction of the field virtually without changing their shape. This shift takes place due to the magnetoplasmonic nonreciprocity effect in accordance to Eqs. (1) and (3). As a result, we obtain a significant enhancement of the T-MOKE, which is defined as the the relative change in the reflected or transmitted light intensity when a medium is magnetized along two opposite directions:

$$\delta = \frac{I(\mathbf{M})-I(-\mathbf{M})}{I(0)} \qquad (4)$$

where *I(M)* and *I(0)* are the intensities of the reflected or transmitted light in the magnetized and non-magnetized states, respectively [88].

Due to the T-MOKE light intensity can be controlled by magnetic field without any polarizers or other additional optical elements. T-MOKE is mostly determined by interface between nonmagnetic and magnetic media and therefore is highly sensitive to the magnetization near the sample surface and can sustain decent values even for ultra-thin films. Moreover, its inverse counterpart is of primary importance in ultrafast magnetic phenomena [36].

The T-MOKE for a bare iron-garnet film is very small ($\delta \sim 10^{-5}$), while for an MPC it reaches $1.5 \cdot 10^{-2}$ as was demonstrated in Ref. [14] [Figs. 4(a) and 4(c)]. Optimization of the MPC structure along with excitation of the hybrid modes – waveguide-plasmon polaritons provided further increase of the T-MOKE up to 15% [44] [Fig. 4(d)].

A next step forward in this direction was made by Chekhov et al. in [83], where additional layer of bismuth iron-garnet was deposited on top of the gold grating [Fig. 4(e)]. In contrast to the traditional Au/garnet MPCs, spectra of the T-MOKE measured in transmission demonstrate rather specific features: a high-quality resonance for the long-range SPP and a broad 60 nm wide resonance for the short-range SPP [Fig. 4(f)].

A sophisticated multilayer structure consisting of an MPC, with a rare-earth iron garnet microresonator layer and a plasmonic grating deposited on top of it, was fabricated and studied in order to combine functionalities of photonic and plasmonic crystals [Fig. (5)] [41]. The plasmonic pattern also allows excitation of hybrid plasmonic-waveguide modes localized in dielectric Bragg mirrors of the MPC or waveguide modes inside the microresonator layer. These modes give rise to additional resonances in the optical spectra of the structure and to the enhancement of the T-MOKE.

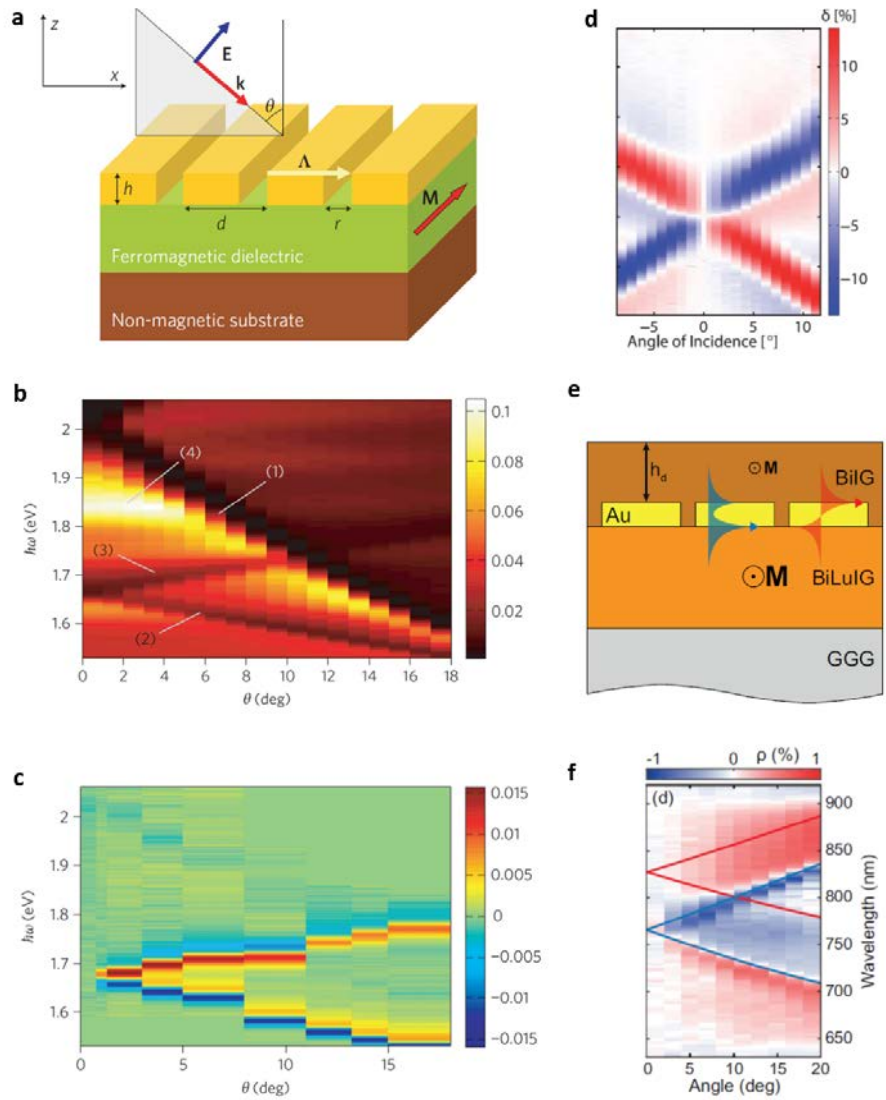

**Fig. 4.** (a) An MPC of a gold grating and a ferromagnetic dielectric (bismuth iron-garnet) illuminated with the incident p-polarized light in T-MOKE configuration. (b-c) False-color plots showing the experimentally measured transmission (b) and the T-MOKE parameter δ (c) as a function of photon energy (vertical axis) and the angle of incidence (horizontal axis). The geometrical parameters are: grating height is $h$ = 120 nm, grating period is $d$ = 595 nm, grating slit width is $r$ = 110 nm. The in-plane magnetic field strength is 200 mT. The features labeled (1) – (4) are related to the SPPs or Fabry-Perot eigenmodes. Reprinted with permission from Ref. [14]. Copyright 2011 Springer-Nature. (d) Measured angle dependence of the T-MOKE in transmission of an MPC similar to that studied in Fig. 4(a)-(c) with an applied external magnetic field of B = 80 mT. For symmetry reasons the effect vanishes for normal incidence and changes sign when the incidence angle is reversed (from 0° to -90°). For angles θ ≥ 4° a remarkably high value of δ = 13% is reached. Reprinted with permission from Ref. [143]. Copyright 2013 Institute of Physics. (e-f) An MPC made of a gold grating placed in between two thin layers of iron-garnet (e), and T-MOKE spectrum observed for such structure (f). Reprinted with permission from Ref. [83]. Copyright 2018 The Optical Society.

T-MOKE in MPCs was shown as an efficient tool for the MO investigation of ultra-thin magnetic films, which allows to access their magnetization state and MO properties [89]. For magnetic films thicker than 40 nm, the T-MOKE marginally depends on the film thickness. A further decrease in the film thickness diminishes T-MOKE since for such thicknesses the SPP field partially penetrates inside the non-magnetic substrate. Nevertheless, the T-MOKE remains measurable even for few-nm-thick films.

Recently magnetoplasmonic quasicrystals have been introduced and demonstrated to provide a unique MO response [90] [Figs. 5(c) and 5(d)]. The quasicrystal consists of a magnetic dielectric film covered by a thin gold layer perforated by slits, forming a Fibonacci-like binary sequence. The T-MOKE acquires controllable multiple plasmon-related resonances, resulting in a MO response over a wide frequency range. Multiband T-MOKE might be valuable for numerous nanophotonic applications, including optical sensing, control of light, all-optical control of magnetization, etc.

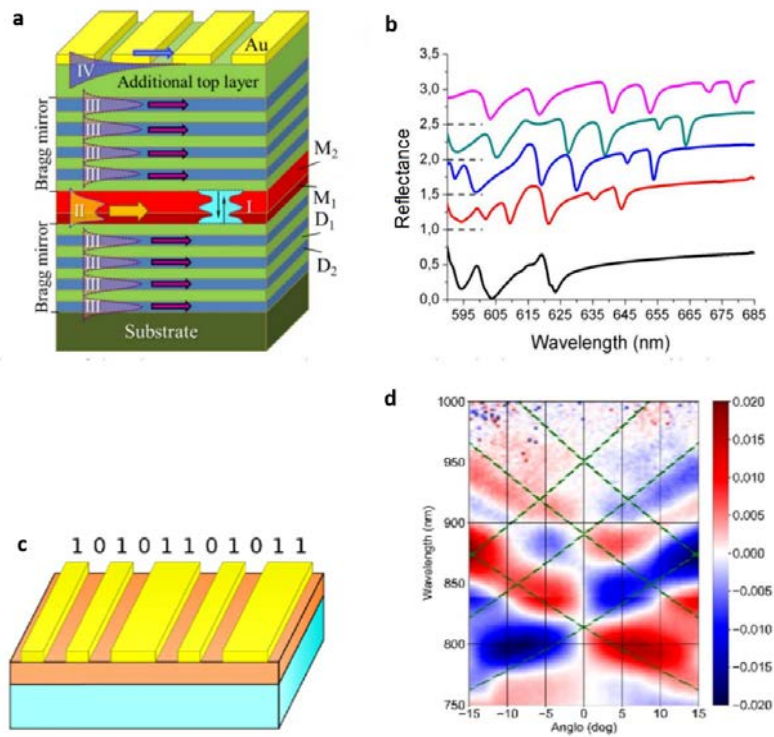

**Fig. 5**. (a) Schematic of the photonic magnetoplasmonic sample. The layers D1 are $SiO_2$ (thicknesses are 117 nm of the stack layers and 100 nm of the top layer), the layers D2 are $TiO_2$ (thickness is 76 nm), layers M1 and M2 are magnetic dielectrics of composition $Bi_{1.0}Y_{0.5}Gd_{1.5}Fe_{4.2}Al_{0.8}O_{12}$ and $Bi_{2.8}Y_{0.2}Fe_5O_{12}$, respectively (the thicknesses are 72 and 271 nm,

respectively), the bars on top depict the gold grating (height $h_{Au}$ = 60 nm, period $d$ = 370 nm, slit width $w_{slit}$ = 220 nm). Filled curves represent schematically the eigenmode profiles in the structure, the arrows show the wavevectors $\beta$ for corresponding modes: a Fabry-Perot microresonator mode (I), a waveguide mode of the microresonator magnetic layer sandwiched between two Bragg mirrors (II), a waveguide mode localized in the higher refractive index layers of the Bragg mirrors (III), a SPP at the gold/dielectric interface (IV). (b) Angular dispersion of the optical resonances experimentally measured in the reflectivity geometry for TM-polarized incident light. Incidence angles (from bottom to top): 0°, 4°, 6°, 8°, 10°. Spectra are shifted vertically by 0.5 arb. unit. Reprinted with permission from Ref. [41]. Copyright 2015 Institute of Physics. (c) Magnetoplasmonic quasicrystal with a thin gold layer perforated by slits, forming a Fibonacci-like binary sequence and (d) its T-MOKE spectra in false-color. Green lines show the calculated dispersion curves for the SPPs. Reprinted with permission from Ref. [90]. Copyright 2018 The Optical Society.

The T-MOKE configuration is of particular interest since it provides also new ways of routing the directivity of light emission by using an external magnetic field. The routing of emission for excitons in a diluted-magnetic-semiconductor quantum well was demonstrated in hybrid plasmonic semiconductor structures [91]. In that case a CdMnTe quantum well sandwiched between a CdMgTe buffer and spacer layers was covered with 1D gold grating to allow SPP excitation which l led to enhanced light emission directionality of up to 60%.

#### d. Magneto-optical effects in longitudinal magnetization of a MPC

As we have discussed so far, the implementation of nanostructured hybrid materials provides a remarkable increase of the T-MOKE. Interestingly, plasmonic structures can give origin to novel MO phenomena as well [25,40] In particular, the plasmonic crystal consisting of a 1D gold grating on top of a magnetic waveguide layer allows observing the MO intensity effect in longitudinal configuration, where a magnetic field is applied in the plane of the magnetic film and perpendicular to the slits in the gold grating [Fig. 6]. Longitudinal magnetization of the structure modifies the field distribution of the optical modes and thus changes the mode excitation conditions. In the optical far-field, this manifests in the alteration of the optical transmittance or reflectance when the structure is magnetized. Thus, this effect is described similarly to the T-MOKE by relative change of the transmittance or reflectance but this time one should compare demagnetized ($T_0$) and longitudinally magnetized ($T_M$) states: $\delta = (T_0-T_M)/T_0$. Such MO response represents a novel class of effects related to the magnetic field-induced modification of the Bloch modes of the periodic hybrid structure. Therefore, we define it as the *longitudinal magnetophotonic intensity effect* (LMPIE).

It follows from the Maxwell's equations with the appropriate boundary conditions, that the two principal modes of the magnetic layer – transverse magnetic (TM) and transverse electric (TE) modes – acquire in the longitudinal magnetic field additional field components, which are linear in gyration *g*, and thus turn into quasi-TM and quasi-TE modes, respectively [25,92] [Fig. 6]. Thus, the coupling between the TE and TM field components emerges in the magnetized layer. This leads to the origin of the LMPIE.

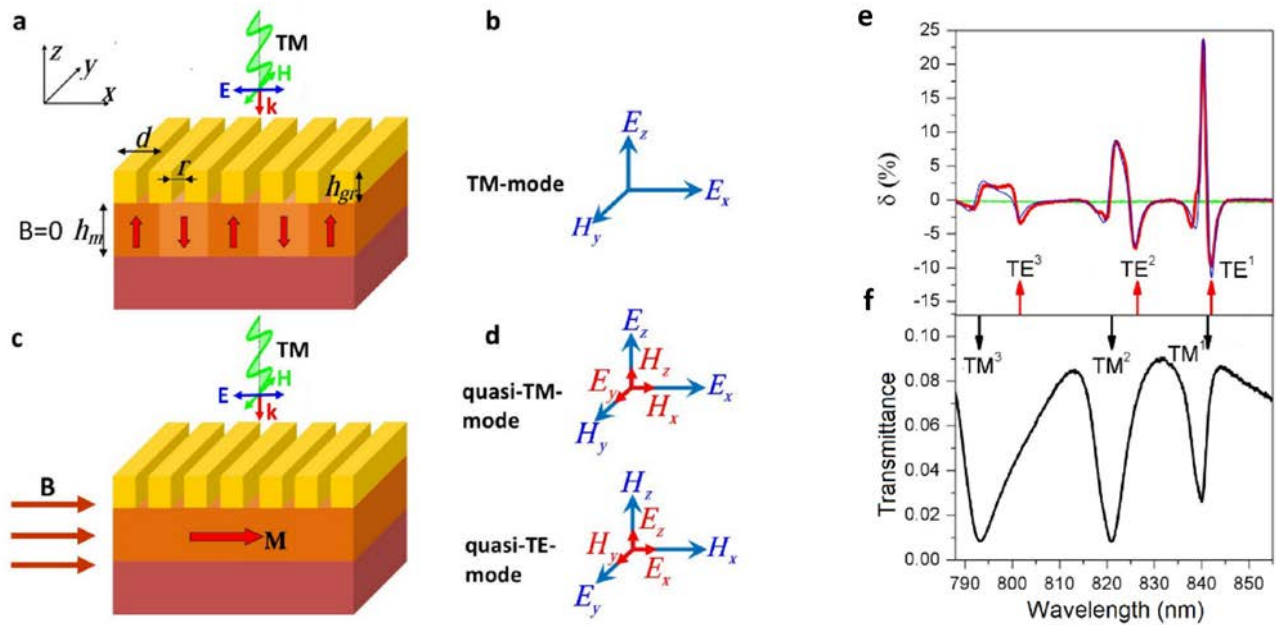

**Fig. 6**. (a, c) Schematics of the MPC studied in Ref. [25] in (a) demagnetized (multidomain) and (c) longitudinally magnetized conditions. The MPC consists of a gold grating of height $h_{gr}$ stacked on a smooth ferromagnetic dielectric of thickness $h_m$ grown on a non-magnetic substrate. The gold grating has period *d* and slit width *r*. (b, d) Optical modes which can be excited by incident TM-polarized light for the (b) demagnetized and (d) longitudinally magnetized structure. The long blue arrows represent the principal field components associated with TM and TE modes in the non-magnetic case, while the short red arrows indicate the components induced by the longitudinal magnetization. Magnetophotonic effect measured via the intensity modulation induced by magnetizing the MPC. (e) Spectrum of the LMPIE when a magnetic field B = 320 mT reaching almost the saturation value is applied. The blue curve shows the calculated values of $\delta$. There is no LMPIE for the bare magnetic film (green curve). (f) Spectrum of the optical transmittance for the demagnetized structure. Black and red arrows indicate calculated spectral positions of the quasi-TM and quasi-TE resonances, respectively. The modes are denoted by the number of their $H_y$ or $E_y$ field maxima along the $z$-axis. The light beam is TM-polarized and it is incident on the sample at normal incidence. The sample parameters are: $d = 661$ nm, $h_{gr} = 67$ nm, $r = 145$ nm, $h_m = 1270$ nm. Reprinted with permission from Ref. [25]. Copyright 2013 Springer-Nature.

Experimental demonstration of the LMPIE was performed for an MPC based again on a bismuth iron-garnet film [25]. A prominent feature of this sample is that it was designed such that the dispersion curves of the principal TM and TE modes correspond to the 2$^{nd}$ diffraction order intersection at the Γ point ($\kappa = 0$) of the Brillouin zone. As a consequence, both modes can be excited by normally incident light at the same frequency. The LMPIE was observed in transmission. No intensity modulation occurs for the bare magnetic film [Fig. 6(e), green curve]. The longitudinally applied magnetic field resonantly increases transmittance by 24% at $\lambda = 840$ nm, where both modes are excited [Figs. 6(e) and 6(f)]. There are also resonances at about 825 nm and 801 nm, corresponding to excitation of the TE-modes only, though their values are several times smaller. The measured magnetophotonic intensity effect with 24% modulation can be considered giant since it is a second-order effect in gyration (as it is an even function of the magnetic field). The modulation level can be increased even further by using materials with higher MO response, thus enabling the use of the LMPIE in modern telecommunication devices. Furthermore, the effect of mode switching is of great interest in the framework of active plasmonics and metamaterials [92,93]. Recently, the LMPIE in an MPC was used for magnetometry [94]. The experimental study revealed that such an approach allows to reach the nT sensitivity level, which was limited by the noise of the laser. Moreover, the sensitivity can be improved up to fT/Hz$^{1/2}$ and micrometer spatial resolution can be reached.

### e. Magneto-optical effects in dot- and antidot periodic arrays

Metallic grating-like structures can provide the basis for the excitation of both SPPs and surface lattice resonances (SLRs) [95-98]. By employing modern nanolithography techniques, 2D grating geometries can be designed at will, thus facilitating further tunability in the launching and control of the momentum of such plasmon modes.

For the case of SPPs, the grating can provide the additional momentum needed for triggering these evanescent surface waves according to the master equation:

$$\vec{k}_{SPP} = \vec{k}_{//} + \vec{G} \qquad (5)$$

with $\vec{k}_{SPP}$ being the SPP wave vector, $\vec{k}_{//}$ the component of the incident light wave vector parallel to the grating plane and $\vec{G}$ a reciprocal lattice vector corresponding to the grating geometry. Furthermore, by performing the transformation into the reciprocal space of the grating geometry, it is possible to adapt the established Ewald geometrical representation, often utilized in scattering studies [99], to discuss the coupling of the incident light to plasmon resonances. However, it is important to note here that, due to the momentum gap between the dispersion relations of light propagating in vacuum and SPPs, the situation resembles that of an "*inelastic*"-like process, where the grating effectively adds the missing momentum to the light, thus enabling the launching of a SPP.

To illustrate the use of this geometric construction and its utility in designing an MPC, we refer to the well-studied case of hexagonal antidot arrays [50,52-54], schematically shown in Fig. 7(a). Different cases of illumination geometry are illustrated in Figs. 7 (d) - 7(g). In adopting this approach, we utilize the transformation into reciprocal space to examine the wave vector relationships for the light, propagating plasmon modes and reciprocal lattices vectors [Figs. 7(b) and 7(c)]. This representation can yield direct insight into the possible momentum amplitudes and directions with respect to the reciprocal lattice that can enable SPP modes. It is also possible to examine the conditions for excitation of collinear or non-collinear plasmon modes with respect to $\vec{k}_{//}$ or the reciprocal lattice vectors, as well as the direction of propagation (back or forward). SPPs in magnetoplasmonic hexagonal antidot structures of Ni were recently imaged using photo-emission electron microscopy [50]. The photo-emitted electron density can be correlated to the local electric field strength, which is modified by the presence of SPPs. This local field intensification leads to an increase of the pure MO contribution in terms of Fresnel reflection coefficients, such as $r_{\text{sp}}$, which are proportional to the electric field inside the MO layer and result in enhanced Kerr rotation [50].

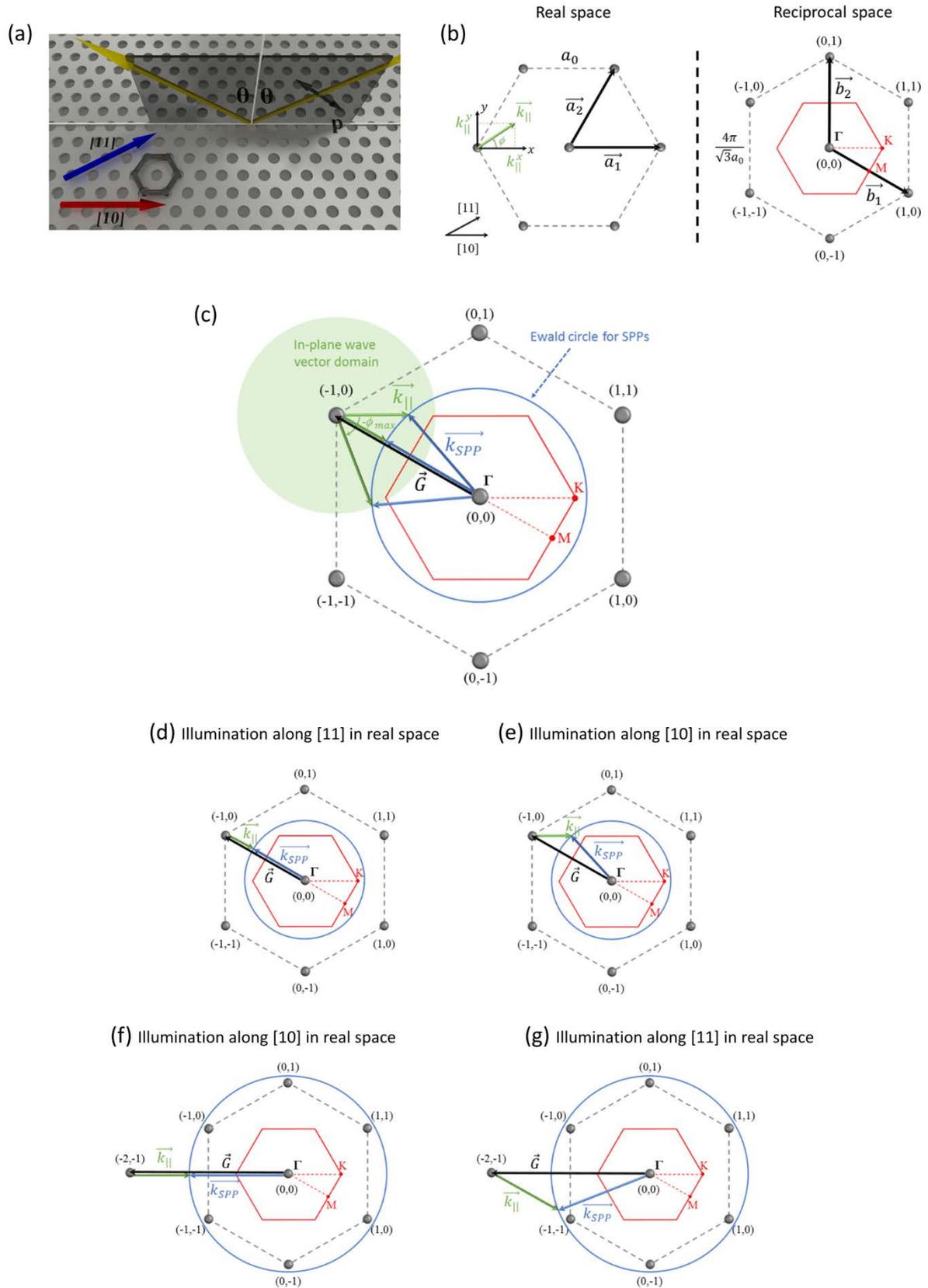

**Fig. 7**. (a) A schematic depiction of a hexagonal antidot sample, resulting from the evaporation of a thin metallic film on a self-organized colloidal shadow mask. The lattice constant $a_0$ can be designed to be in the submicron range. Incident light at an angle $\theta$ from the normal and bearing a certain polarization (p polarization shown here) can be used to launch

SPPs. (b) The real and reciprocal space lattices of the hexagonal antidot structure depicted in (a), with the respective unit vectors. (c) The geometrical construction scenario for $\vec{G}_{(-1,0)}$ vector, highlighting the matching conditions for a given incident light wavelength. All the points on the Ewald circle for SPPs (blue) contained within the in-plane vector domain (green disc, defined by all possible $(\theta, \phi)$ pairs) can be excited. (d) The case for illumination with the scattering plane along the [11] real space direction. For certain illumination conditions (wavelength, angle of incidence) a reciprocal lattice vector can be matched ($\vec{G}_{(-1,0)}$ shown here), and an SPP is launched with $k_{SPP} > k_{//}$. All momentum vectors are collinear in this case. (e) The case for illumination with the scattering plane along the [10] real space direction and for matching the $\vec{G}_{(-1,0)}$ (similar to (d)) reciprocal lattice vector. The momentum vectors are non-collinear now. (f) The case for illumination with the scattering plane along the [10] real space direction and for matching the $\vec{G}_{(-2,-1)}$ reciprocal lattice vector. The momentum vectors are collinear. (g) The case for illumination with the scattering plane along the [11] real space direction and for matching the $\vec{G}_{(-2,-1)}$ (similar to (f)) reciprocal lattice vector. The momentum vectors are non-collinear again.

In contrast, a periodic arrangement of metallic particles can also result in sharp resonances, which are referred to as SLRs in literature [58,59,97,100-102]. This phenomenon arises from the diffracted light propagating in the plane of the sample, strongly coupling to localized plasmon resonances of individual particles, collectively resulting in a dramatic narrowing of the plasmon resonances. Whenever such a condition is met by tuning the angle of incidence and wavelength of the light, abrupt changes in the reflectivity are observed, which are commonly referred to as Wood's anomalies [103]. If the metallic particle array is composed of ferromagnetic materials, the Wood's anomalies are accompanied by an enhancement of the MOKE, in a similar fashion to the case of SPPs [58,59], as shown in Fig. 8.

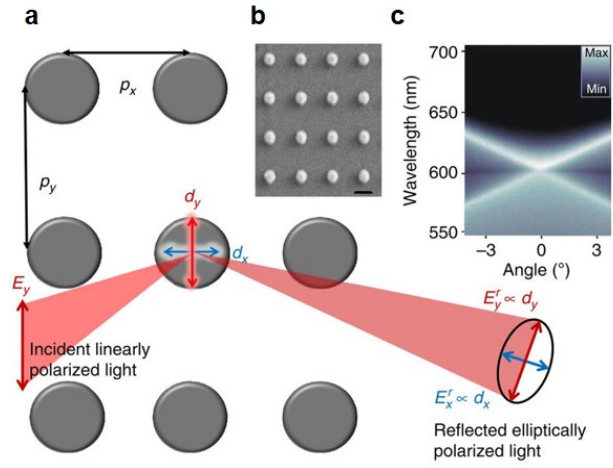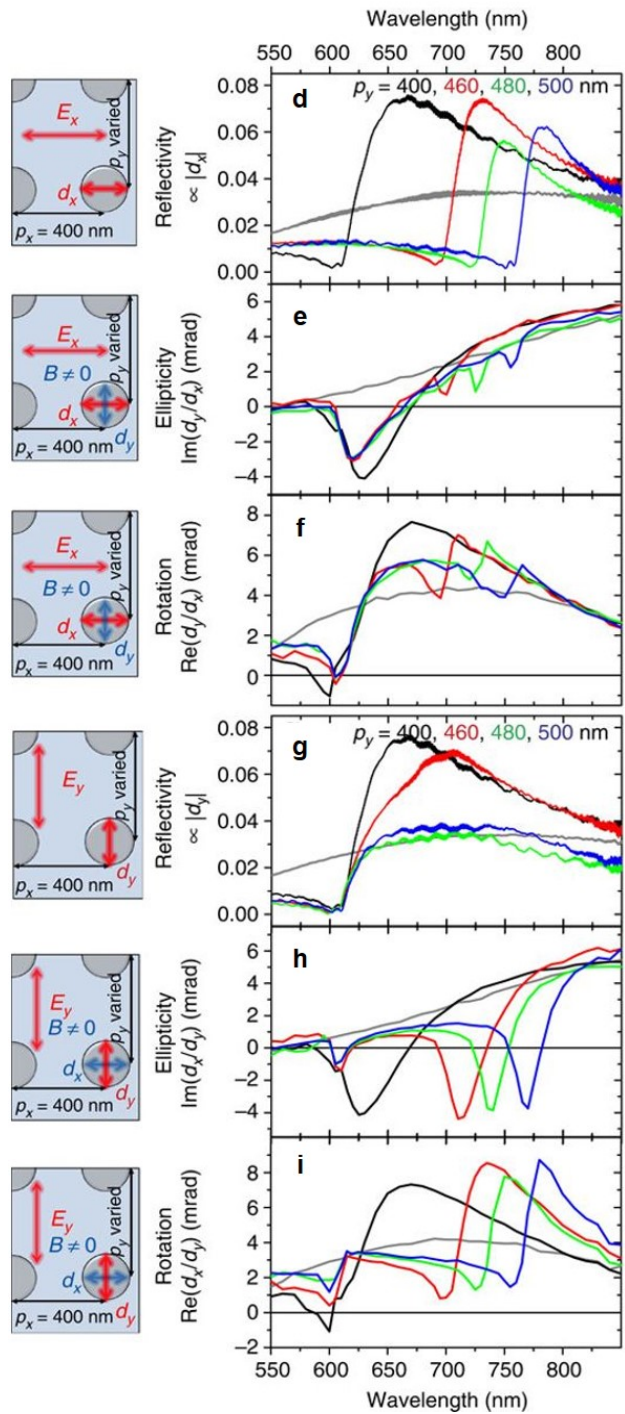

**Fig. 8.** (a) Schematic of the system studied in ref [59]. In the presence of magnetic material, the system response is governed not only by the induced dipole moment $d_y$ parallel to the driving field $E_y$ and the lattice period $p_x$ (direction of dipole radiation), but also by the spin–orbit-induced and magnetic-field tunable dipole moment $d_x$ and lattice period $p_y$. (b) Scanning electron micrograph of an ordered rectangular array of cylindrical Ni nanoparticles. Scale bar, 200 nm. (c) Angle- and wavelength-resolved optical transmission of a sample with $p_x=p_y=400$ nm and with particle diameter 120 nm, showing crossing of the <+1, 0> and <−1, 0> diffracted orders of the lattice at normal incidence. Normal incidence experimental optical reflectivity (d), MO Kerr ellipticity (e) and rotation (f) with polarization $E_x$. Normal incidence experimental optical reflectivity (g), MO Kerr ellipticity (h) and rotation (i) with polarization $E_y$. The black, red, green and blue curves correspond to the periodicities $p_y = 400, 460, 480$ and 500 nm, in all the figures. The grey line corresponds to a random array of Ni nanodisks. Reprinted with permission from ref [59]. Copyright 2015 Springer-Nature.

In an analogous manner, the propagation direction of the diffracted beam of order *n* is determined by application of the Laue condition, accounting for the in-plane (*x*) projections of all wave vectors

$$k_x^{out} = k_x^{inc} + nG \quad (6)$$

with $k_x^{inc} = k_0 \sin\theta$, $k_0 = 2\pi/\lambda$, and $G = 2\pi/\Lambda$, where $k_0$ is the wave vector of the incident light, $\theta$ the angle of incidence, $\lambda$ the wavelength and $\Lambda$ the grating period [see also Fig. 9]. It is worth noting here that, compared to the SPP case, this case resembles an "*elastic*"-like process. Therefore, the momentum added by the grating is used to change the direction of the light wave vector, while the length of the latter is the same for the incident and scattered light. Introducing magnetic metallic particles with anisotropic shapes further allows distinct and anisotropic in-plane SLRs, determined by the particle polarizabilities and the spectral relation between localized resonances and Bragg modes [58]. Such MPCs could yield new metamaterials and optical devices, like magnetism-controlled nonreciprocal optical isolators, notch-filters for the light polarization, or bio-sensors [58].

An interesting situation arises when the grating period $\Lambda$ is such that $\lambda/4 < \Lambda < \lambda/2$ (or $2k_0 < G < 4k_0$), where in the non-linear optical regime the sample can be treated as a grating structure, while in the linear regime the sample behaves as a metasurface [102,104,105]. This situation is graphically depicted in Fig. 9, for the case of a sample composed of Ni nanodimers, having different periodicities along the *x* and *y* directions, defining grating- and metasurface-like behavior in the nonlinear and linear response, respectively. The second harmonic generation exhibits a grating behavior, with associated Wood's anomalies. A decrease in specularly reflected intensity, of an order of magnitude larger than the linear effect (in the grating regime) is observed, accompanied by a sizeable magnetic contrast [102].

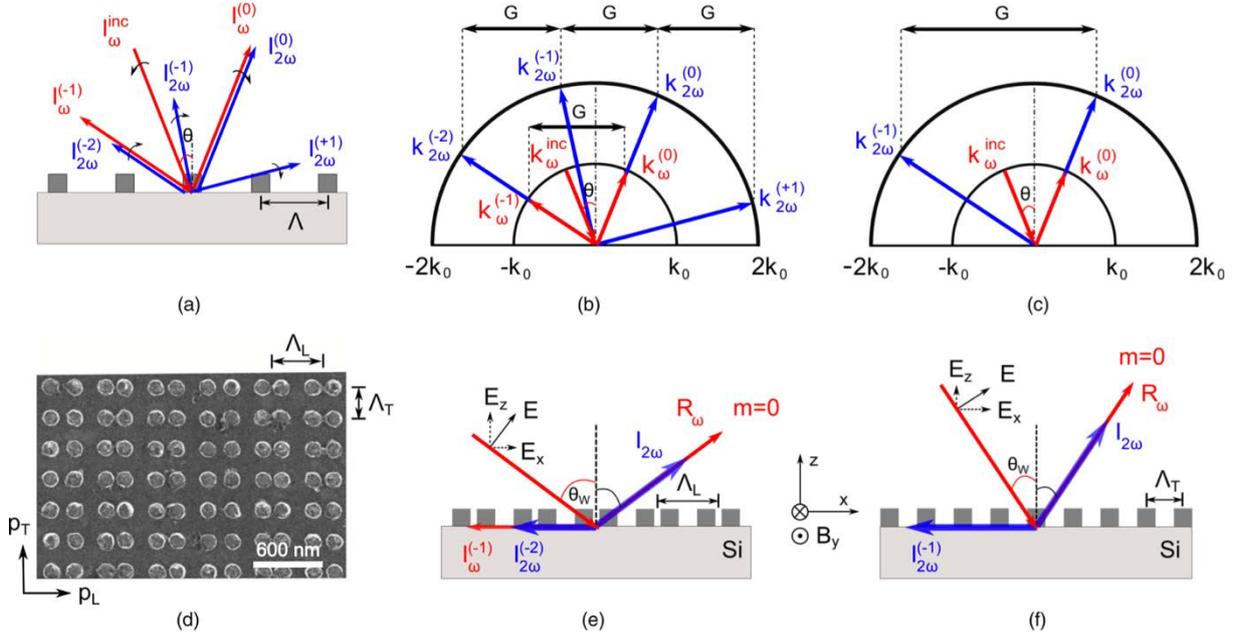

**Fig. 9**. (a) Diffraction from a grating with $\Lambda > \lambda/2$, shown for both linear and non-linear second harmonic diffracted beams (red and blue, respectively). (b) Depiction of the diffraction in the grating regime, where $G \leq 2k_0$ or $\Lambda \geq \lambda/2$. All diffracted vectors move on circles with radii defined by their corresponding regime order ($k_0$ for the fundamental and $2k_0$ for the second harmonic) and are separated by a multiple of the reciprocal lattice vector $\vec{G}$, reflecting the generalized Laue condition between the in-plane projections. (c) Transition regime, with $2k_0 < G < 4k_0$ or $\lambda/4 < \Lambda < \lambda/2$. The sample acts as a grating for the second harmonic and as metasurface in the fundamental. (d) Scanning electron microscopy image of a rectangular arrays of Ni nanodimers. (e) Excitation geometry for the Wood anomaly in the grating regime and at an angle of incidence $\theta=\theta_w$. (f) The pure non-linear Wood anomaly in the metasurfaces regime for the fundamental, where only a non-linear diffracted beam emerges at $\theta=\theta_w$. Reprinted with permission from Ref. [102].

# III. Applications of nanoscale magnetoplasmonics and magnetophotonics

### a. Biological and chemical sensing

A remarkable property of nanostructured metal systems is their ability to concentrate the optical energy on a nanoscale volume. Plasmons (propagating and localized) are strongly confined at the interface between two media with permittivities of opposite sign, such as the interface between a dielectric and a metal. When the incident radiation couples to such plasmon modes, clear signatures, viz. plasmon resonances, in the optical response of the system are observed. The operating principle of plasmon-based sensors is based on the registration of the spectral and angular positions of these resonances, which depend strongly on the optical properties of the surrounding medium, such as, for instance, its refractive index [Fig. 10]. This is why plasmonic nanostructures are often used as transducers for single molecular recognition and for gas sensing applications [106-108]. Due to their sharp LSPRs or SPPs resonances, noble metals are the preferred materials to build such sensors, although recently it has been proved quite widely that a remarkable exception is the application of magnetoplasmonic structures in label-free molecular detection, bio-chemo-sensing, determination of nanoscale distances. In these cases, despite the insignificant MO activity and large absorption losses, MPCs and nanoantennas were found to enable a radically improved sensitivity, clearly outperforming conventional plasmon based sensors and rulers. In MO-active systems, the excitation of the plasmonic modes affects also their MO response. Furthermore, the enhancement of the sensor sensitivity (which is the derivative of the acquired measurement with respect to the analyzed refractive index change) and the resolution limit, as well as reliability and reproducibility, remain essential. A way to improve the sensitivity of nanostructures can be represented by the use of magnetoplasmonic nanostructures instead of the classical noble metal-based nanosensors. In this case, rather than measuring the usual transmission or reflection signals, tracking of the MO response can lead to a real sensitivity advantage when compared to absorbance-based measurements. As such, magnetoplasmonic nanostructures could form the basis of highly sensitive label-free biosensors. In 2006, Sepulveda et al. [109] showed an increase in the limit of detection by a factor of 3 in changes of the refractive index and in the adsorption of biomolecules compared with regular plasmon sensors. Years later, Regatos et al. presented an improvement of the previous approach, demonstrating a twofold increase in the MO-based sensors' detection limit with respect to the intensity-interrogated SPPs-based biosensors operating with refractive-index changes [110]. In 2013, Martín-Becerra et al. showed that the modulation of the SPP wavevector induced by an external applied magnetic field represents a new parameter with a higher sensitivity to the refractive index than the SPP wavevector, so monitoring it can lead to sensors with improved properties [111]. In 2014, Manera et al. demonstrated that magnetoplasmonic sensors can be used for investigating biomolecular interactions in liquid phase with higher sensing performance in terms of sensitivity and lower limit of detection with respect to traditional SPR sensors [112].

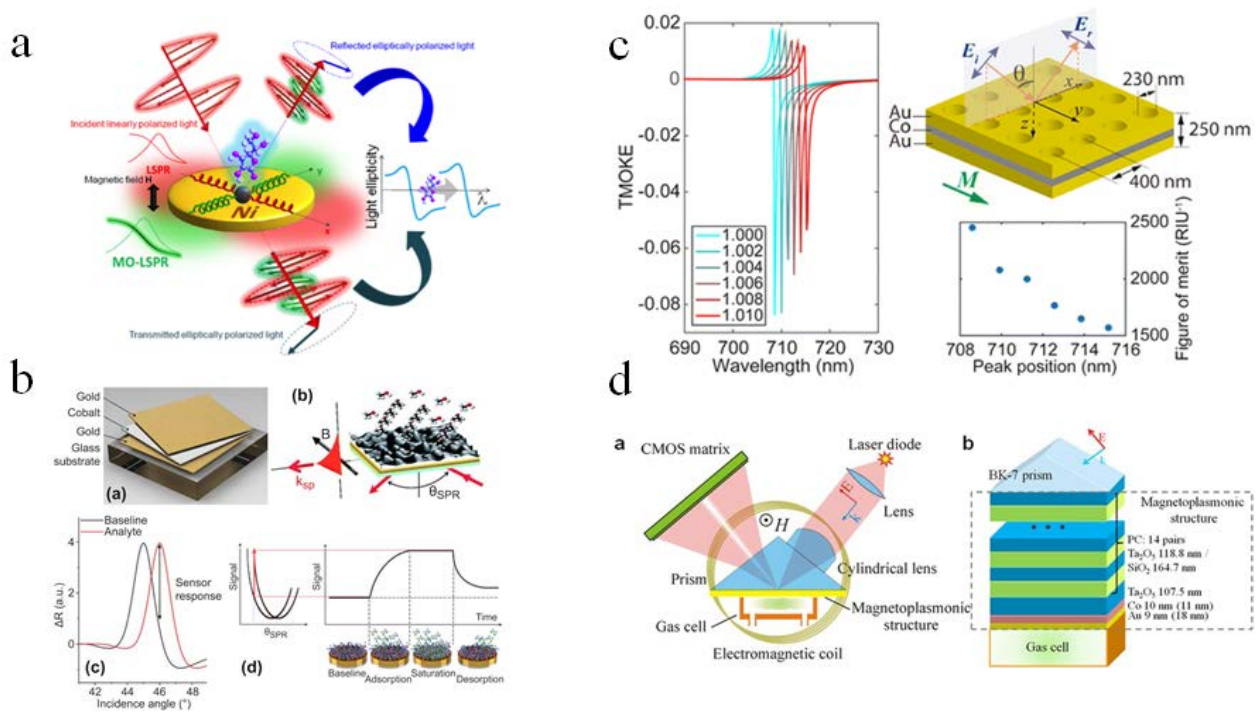

**Fig. 10.** (a) Light polarization manipulation enabled by phase compensation in the electric response of a magnetoplasmonic nanoantenna controlled through a precise design of the LSPR induced by the MO activity (MO-LSPR) of the ferromagnetic constituent material (Ni) and exploitation of the effect for ultrasensitive molecular sensing. Reprinted with permission from Ref. [123]. Copyright 2015 Springer-Nature. (b) Upper-left panel: illustration of the composition

of magnetoplasmonic multilayers (composed of noble and ferromagnetic metals). Upper-right panel: upon SPP excitation, an enhancement in the T-MOKE signal, ΔR, is recorded. The direction of the applied magnetic field (B), with respect to the incident plane, is indicated by the black arrow. $\theta_{SPP}$: incidence angle of the light that causes SPP excitation. Bottom-left panel: the enhanced T-MOKE signal makes it possible to sensitively probe refractive index changes at the metal/dielectric interface of MO-SPP transducers. Bottom-right panel: illustration of the different stages of the sensing process; a.u.: arbitrary units. Reprinted with permission from Ref. [113]. Copyright 2016 SPIE. (c) Theoretical structure introduced by Caballero et al. showing an ultrahigh figure of merit. Reprinted with permission from Ref. [79]. Copyright 2016 American Chemical Society. (d) Magnetoplasmonic sensing structure for gas detection proposed and experimentally realized by Ignatyeva et al. with a quality factor > 1000. Reprinted with permission from Ref. [117]. Copyright 2016 Springer-Nature.

Manera and co-workers used hybrid Au/Co/Au magnetoplasmonic nanoparticles as novel transducer probes to achieve enhanced sensitivity in SPP-based chemosensors [113] [Fig. 10(b)]. Similar systems supporting SPPs and MO functionalities with improved sensitivities were then proposed by other groups [114-116]. More recently, Ignatyeva et al. presented a novel concept of magnetoplasmonic sensor with ultranarrow resonances and high sensitivity, with a quality factor exceeding 1000 [117] [Fig. 10(d)]. In another work, they showed that high-quality factor surface modes in photonic crystal/iron-garnet film heterostructures can be used also for sensor applications [118]. In 2016, Caballero et al. proposed Au–Co–Au films perforated with a periodic array of subwavelength holes as transducers in MO-SPP sensors, introducing a detection scheme showing figures of merit that are 2 orders of magnitude larger than those of any other type of plasmonic sensor [79] [Fig. 10(c)]. A similar structure has been later proposed also by Diaz-Valencia et al. [119].

All the previously mentioned systems were based on the excitation of propagating plasmons. Until 2011, nobody was thinking to use magnetoplasmonic systems supporting LSPRs as sensors, when Bonanni et al. showed that nickel nanoantennas can be used as highly sensitive detectors of refractive index changes [51]. In 2012, Zhang and Wang proposed magnetoplasmonic FePt-Au nanorods as a novel type of nano-bioprobe that allow simultaneous magnetic manipulation and optical imaging for single molecule measurements, drug delivery, in vitro and in vivo diagnostics and therapy [120]. Later, Pineider et al. used gold nanoparticles dispersed in liquid solvents for refractometric sensing by magnetic circular dichroism (MCD) measurements [121]. Similarly, Manera et al. showed recently that monitoring the MO response of gold nanoantennas measured in Kretschmann configuration has the practical advantage to pursuit a better signal-to-noise ratio, an essential requirement for high resolution sensing signals [122]. In 2015, Maccaferri et al. brought up a novel application for magnetoplasmonic antennas made of pure nickel, suggesting to monitor the polarization ellipticity variation of the transmitted and reflected light, showing a raw surface sensitivity (that is, without applying fitting procedures) corresponding to a mass of less than 1 attogram per nanoantenna of polyamide-based molecules, which are representative for a large variety of polymers, peptides and proteins [123] [Fig. 10(a)]. A similar approach has been also used recently by Tang. et al. to increase the performances of refractive index sensors by magnetoplasmons in nanogratings [124], and also by other groups implementing the method of tracking the ellipticity (and rotation) of the transmitted or reflected light using chiroptical nanostructures [125] and pure plasmonic nanoantennas [126]. In 2015, Zubritskaya et al. demonstrated that nickel dimers are able to report nanoscale distances while optimizing their own spatial orientation with about 2 orders of magnitude higher precision than current state-of-the-art plasmon rulers [66]. In 2016 Herreño-Fierro et al. showed that ellipsometric phase-based transducers can be used for bio-chemical sensing purposes [127]. In 2018, Pourjamal et al. showed that hybrid Ni/SiO$_2$/Au dimer arrays display improved sensing performances if compared to random distributions of pure Ni nanodisks and/or their random counterpart [128]. Finally, it is worth mentioning that, depending on the specific application, it could be more appropriate to exploit systems supporting either SPPs or LSPRs. Specific binding events can be detected using either SPPs [113] and LSPRs [126], but LSPRs or localized modes in general have been proved to be superior in molecular sensing of single molecules [129].

**b. Light polarization and phase control**

The investigation of the phenomena arising from the mutual interplay of magnetism and light-matter interactions in spatially confined geometries is decisive for data communication, photonic integrated circuits, sensing, all-optical magnetic data storage and light detection and ranging. Light polarization rotators and nonreciprocal optical isolators are essential building blocks in these technologies. These macroscopic passive devices are commonly based on the MO Faraday and Kerr effects. Magnetoplasmonic nanoantennas and MPCs enable the magnetic control of the non-reciprocal light propagation and thus offer a promising route to bring these devices to the nanoscale, featuring dynamic tunability of light's polarization and phase. Because noble metals provide outstanding light localization and focusing, combining them with ferromagnets became a wise strategy in design of active magnetoplasmonic devices operating with SPP [25,28-30,32-42,44]. Similar approach was later utilized with nanostructures supporting localized plasmons. Magnetic field-dependent modulation of the polarization of reflected/transmitted light (magneto-optic Kerr/Faraday effects), owing to the intertwined plasmonic and MO properties, has been reported in Au/Co/Au multilayered [46,130] nanoantennas. It was long believed that ferromagnetic nanostructures alone can not support localized plasmons due to high damping. The discovery of plasmons and their near-field mapping in nickel [48] opened up a promising route of magnetoplasmonic devices for light polarization control, now also in transmission due to their transparent nature. In 2011 Bonanni et al. discovered strong magneto-plasmon localization and plasmon-induced enhancement and tunability of MO activity in pure Ni nanoantennas [51] [Fig. 11(a)]. Similar effects were also demonstrated along the years in nano-perforated/corrugated ferromagnetic films [49,50,52,53,54]. In 2013 Maccaferri et al. developed the analytical model that highlighted the role of SO coupling on MO response [56]. In 2014, this work was followed by the derivation of MO design rules for engineering of nanoantennas with tailored MO response in all Kerr effect configurations (longitudinal, polar and transverse) by Lodewijks et al. [60]. Giant enhancement of MO effects was also observed in hybrid ferromagnetic/noble metal films (MPCs) by Chin et al.

[Fig. 11(c)] [23], Caballero et al. [79] and Kalish et al. [92], graphene/noble metal nanowires [131], and in a ferromagnetic metal/dielectric nanoparticle system where non-metallic high-refractive index semiconductor ('all-dielectric') nanoantennas support optical Mie resonances [132]. Theoretical systems of pure all-dielectric and ferromagnetic nanoantennas with strongly enhanced MO has been proposed [133]. Furthermore, another intriguing and interesting advance in the field is also represented by the manipulation of structured light, namely light carrying orbital and/or spin angular momentum information [134-136]. In these works, either the spin or the orbital angular momenta were shown to be actively tuned by applying an external magnetic field. In particular, an interesting direction might be the merging of the strong chiral response and the angular momentum selectivity reported in Ref. [135] with the strong magnetic field modulation (beyond 100%) reported in [136], where on the contrary the overall chiral response was very weak due to the 2D geometry of the system. Along this direction, it is worth mentioning that a detailed analysis which might help to reach this goal was reported by Feng et al. [137], where they analyzed the contributions from optical chirality, optical anisotropy and magnetic modulation of circular dichroism (CD) to the global optical response in Au/Co split-ring geometries. In this particlaur case they showed a system which have a strong chiral response with a MO-mediated magnetic modulation of CD of about 25% [Fig. 11(d)].

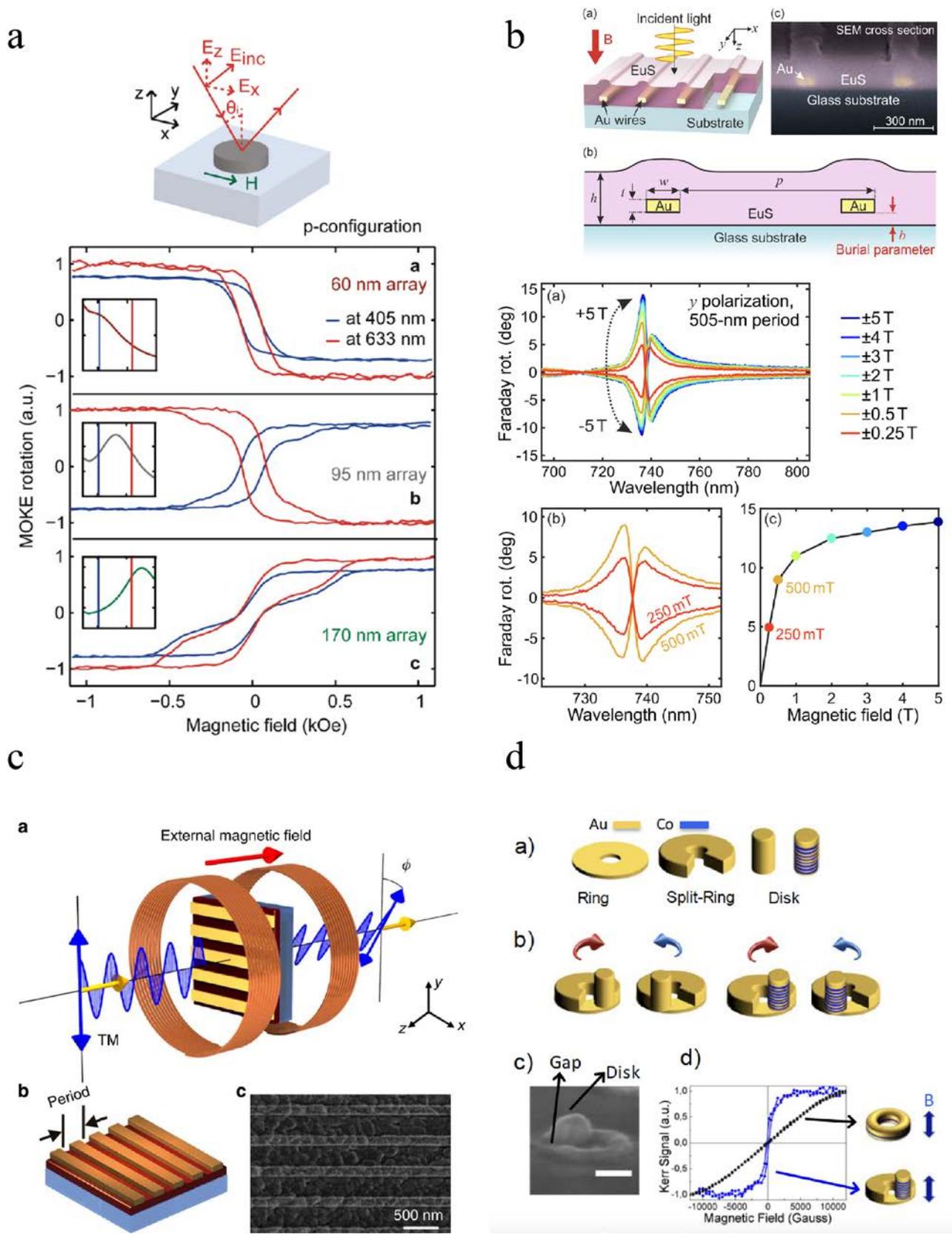

**Fig. 11.** Panel (a): Normalized p-polarization MOKE on nickel nanodisks of 60 (a), 95 (b), and 170 nm (c), employing two different excitation wavelengths (405 nm (graphs in blue) and 633 nm (graphs in red)). The sketch on top is a schematic of L-MOKE in p-configuration. The hysteresis loops measured at 405 nm have been scaled by 80% for clarity of presentation. The insets schematically show far-field extinction spectra for the corresponding nanodisk types in relation to the two excitation wavelengths of MOKE experiments. Reprinted with permission from Ref. [51]. Copyright 2011 Americal Chemical Society. Panel (b), top: (a), (b) Schematic drawing of the sample geometry. p, nanowire period; t, w, gold nanowire thickness and width, respectively. The wires are buried with a distance $b$ between the glass substrate and their lower edge. The nominal thickness of the EuS MO waveguide is $h$, which is increased near the position of the gold nanowires. (c) Colorized SEM micrograph of the sample cross section. The samples are measured at T = 20 K. Panel (b), bottom: (a) Magnetic field dependence of the polarization rotation for a period of 505 nm. (b) Closer view on the rotation spectra for weak magnetic fields. Already for 250 mT the Faraday rotation reaches values of over 4°. (c) Saturation behavior of the Faraday rotation at 737 nm. Reprinted with permission from Ref. [147]. Copyright 2017 Americal Physical Society. Panel (c): (a) Faraday rotation of an MPC for TM-polarized incident light, where φ is the Faraday rotation angle. At normal incidence, TM-polarized light has the electric field perpendicular to the gold wires, and TE-polarized light has the electric field parallel to the wires. (b) Schematic of the MPC, where BIG film (dark red) is deposited on glass substrate (blue) and periodic gold nanowires (golden) are sitting on top. (c) A SEM image of one sample. Reprinted with permission from Ref. [23]. Copyright 2013 Springer-Nature. Panel (d): (a) The different building blocks that will form part of the final structures. (b) Resulted chiral plasmonic and magnetoplasmonic metastructures fabricated and studied Ref [137]. The specific location of the Au or Au/Co multilayer disk at either side of the split ring edge induces an in plane optical anisotropy and determines the handedness of the system. (c) SEM image of a representative structure obtained in this way. Scale bar: 100nm. (d) Comparative MO hysteresis loops for an Au/Co/Au ring structure with in plane magnetic anisotropy and a split ring/ring structure with an Au/Co multilayer, with perpendicular magnetic anisotropy. As it can be seen, magnetic saturation along surface normal requires a much smaller magnetic field for the multilayer case. Reprinted with permission from Ref. [137]. Copyright 2017 The Optical Society.

The ability to externally control optical states could be a key feature in such nanophotonic applications as nanoscale local polarization detection, chirality recognition and polarization spectroscopy, as well as magnetic field sensing [94] or tunable near-field emission of a desired optical state [138]. MO properties of CoPt nanostructures with antiparallel magnetic alignment combined with noble metal (Au and Ag) fine grains were recently investigated by Yamane et al. [139] revealing the enhancement of MO effects via LSPRs in the grains. Previously, the same group achieved an impressive rotation of 20° in the visible spectral region by using CoPt/ZnO/Ag multilayered structure that works like a MO Fabry–Pérot cavity [140]. Almpanis et al. predicted also similar impressive values of the polarization rotation in the near-IR in magnetic garnet film sandwiched between two metallic layers, patterned with periodically spaced parallel grooves on their outer sides [141]. An intriguing case of magnetic field-assisted dynamic alignment resulting in enhancement/cancellation of plasmon optical response rather than polarization was demonstrated utilizing multisegmented Au/Ni/Au nanorods [24]. The recent comparison study on hybrid magnetoplasmonic gold-magnetite nanoparticles with core-shell, dumbbell-like and cross-linked geometries suggests the improvement of tunability, light scattering enhancement and local field enhancement at the interface between magnetic and plasmonic constituents [142].

It is worth mentioning that a magnetic manipulation of propagating plasmons in MPCs made of magnetic garnets [27] can also lead to strongly enhanced MO activity which gives rise to exotic optical properties such as MO transparency [143] and extraordinary transmission in sub-wavelength nanohole arrays [144]. In similar garnet materials, Subkhangulov et al. suggested recently a novel concept for ultrafast MO polarization modulation using terbium gallium garnet ($Tb_3Ga_5O_{12}$), where MO modulation with frequencies up to 1.1 THz is continuously tunable by means of an external magnetic field [145]. Finally, in 2016 Firby et al. proposed a magnetoplasmonic Faraday rotator by incorporating Bi:YIG into a hybrid ridge–plasmonic waveguide structure, which seems to overcome the phase-matching limitations between photonic TE and plasmonic TM modes, thus inducing a 99.4% polarization conversion within a length of 830 μm [146]. Similarly, in 2017 Floess et al. realized a hybrid magnetoplasmonic thin film structure that in transmission geometry displays a giant Faraday rotation of over 14° for a thickness of less than 200 nm and a magnetic field of 5 T at T=20 K [147] [Fig. 11(b)].

The SPPs and the waveguide modes of smooth semiconductors in the presence of an external magnetic field were considered in [148,149]. In these works, Kushwaha and Halevi have undertaken a theoretical study of magnetoplasma waves in a thin, semiconducting film, and they showed that the magnetic field does not introduce any linear magnetization terms in the modes dispersion but it induces transverse electromagnetic field components and the appearance of modes with a negative group velocity, which are a magnetoplasma generalization of the Fuchs-Kliewer modes.

The polarization rotation MO effects were studied in different types of smooth multilayered metal/dielectric structures with either metallic or dielectric magnetized components [147,150-153,176]. Probably, one of the first experimental demonstration of the influence of the plasmonic modes on the Faraday effect was published in [154]. Without making reference to surface plasma waves, author of [154] reported an optically enhanced Kerr rotation in thin iron films, magnetized in the longitudinal orientation, near what has become identified as the plasmon angle.

In some papers [152,153], plasmon-induced P- or L-MOKE enhancement was claimed but it was usually accompanied by decrease in the intensity of the signal. The SPPs-assisted pronounced increase of the Faraday effect was reported in the Bi-substituted iron garnet film covered with thin corrugated silver and gold layers [151]. It was assumed that the main contribution in the enhancement of the Faraday effect in such systems is made by the polarization rotation of the SPPs excited on the metal/dielectric interface.

Faraday and Kerr effects in periodic metal-dielectric structures were also considered recently [70,72,74,144,155-158]. In particular, Diwekar et al. [155] experimentally investigated the Kerr effect upon reflection of visible light from a perforated cobalt film magnetized perpendicular to the surface. It was revealed that, in the vicinity of the region of anomalous transmission of light, the Kerr effect is reduced by one order of magnitude. There is a number of works dealing with the metal-dielectric structures characterized by a considerable enhancement of the Faraday effect [156,157,158]. In those works, a magnetic medium was placed either inside holes in the metal [156], or the metal itself was ferromagnetic [157,158].

The plasmonic crystals of perforated gold on top of the smooth thick ferromagnetic layer were also investigated by measuring the cross-polarized transmission and polar Kerr rotation as a function of external magnetic field [144]. Although the effects of plasmons on these processes were observed, the enhancement of the MO effects via SPPs was not clearly demonstrated.

Though most of the periodic structures were fabricated by means of electron beam lithography and subsequent etching some other fabrication approaches were also used. Sapozhnikov et al. fabricated a 2D MPC by sputtering Co or Ni on top of a PMMA colloidal crystal. It was found that there are some resonance peculiarities in the Kerr rotation spectra. They were attributed to the SPPs and to the resonances related to the multiple interference reflections between the colloidal crystal substrate and the nanostructured film [**77**]. It was found that there are some resonance peculiarities in the Kerr rotation spectra. They were attributed to the SPPs resonances and to the resonances related to the multiple reflections from the interference between reflections from the colloidal crystal substrate and from the nanostructured film. Torrado et al. [**Error! Bookmark not defined.**] also use self-assembling method. They prepared a plasmonic crystal from a polymeric monolayers replicated on nickel. The SPP-assisted increase of the polar and transverse Kerr effects due to the excitation of Ni SPPs modes is reported. However, the effect of disorder was shown to decrease the amount of that enhancement. One more magnetoplasmonic periodic structure was fabricated by depositing Co/Pt multilayers on arrays of polystyrene spheres [159].

It should be noted that the increase of the Faraday and Kerr rotation was reported recently for pure dielectric systems at the wavelengths of waveguide mode resonances [160,161], and in plasmonic structures containing graphene in THz frequency range [162].

In what follows we focus on the Faraday effect in MPC based on iron-garnet films. At the non-resonant frequencies, the Faraday rotation is close to that of a single magnetic film and is defined by the film's thickness. At the eigenmode's excitation wavelength, the resonant features of the MO response is expected. The eigenmodes that are essential for MO behavior are the SPP at the metal/dielectric interface and the waveguide modes of the dielectric layer.

The Faraday rotation can be considered qualitatively as a result of the conversion of the TE-TM field components. Two mechanisms inducing a resonant behavior of the Faraday effect are possible.

Let the incident wave be TM-polarized. First, at the frequency $\omega_{TM}$, where either a TM mode or a SPP can be excited, the TM field is partly converted in a TE mode. But, since the excitation condition for the TE mode is not fulfilled at this frequency, the TE field component is re-emitted contributing to the far field. Moreover, the enhancement of the Faraday effect is due to the fact that the effective path of either the TM mode or the SPP is larger than in the nonresonant case. Second, at the frequency $\omega_{TE}$, the electromagnetic field re-emitted by the structure is partially converted in the TE mode. Also, at this frequency, the TE mode has a large effective path that causes the enhancement of the Faraday effect. Thus, the mechanism for the Faraday rotation enhancement depends on the type of the excited eigenmode.

If the magnetic film thickness is comparable to wavelength of the incoming light, the waveguide modes become essential [30]. As shown in Fig. 12(a), the Faraday rotation displays both negative and positive peaks. Furthermore, the positive Faraday rotation peak at $\lambda = 883$ nm corresponds to more than four-fold enhancement compared to the signal of a continuous magnetic layer of the same thickness. In addition, the positive Faraday rotation peak coincides with the resonance in transmission, allowing about 40% of the incident energy flux to be transmitted. At the same time, the negative Faraday maximum at $\lambda = 818$ nm corresponds to almost negligible transmission. The peaks of the Faraday rotation are also accompanied by abrupt changes in the light's ellipticity. However, the ellipticity becomes zero at the resonance wavelength and the transmitted light remains linearly polarized, but with substantial rotation of the polarization plane.

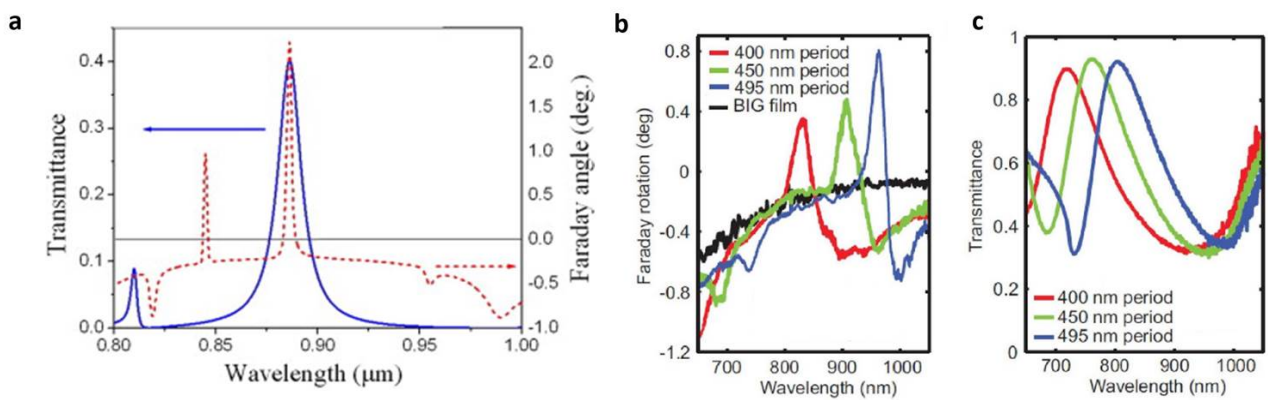

**Fig. 12.** (a) Spectra of the optical transmittance (blue solid line), the Faraday rotation (dashed red line) of an MPC made of an Au grating (lattice period is 750 nm, and bar width 75 nm) of thickness 65 nm and uniform bismuth iron-garnet film of thickness 535 nm. The dielectric is magnetized perpendicular to the sample plane. Reprinted with permission from Ref. [30]. Copyright 2007 Elsevier. Faraday rotation (b) and transmittance (c) of the MPC studied in Ref. [23]. (b) Faraday rotation of the three samples measured at normal incidence (TM polarization), compared with the Faraday rotation of the bare bismuth iron-garnet film. (c) Transmittance of the three samples measured at the normal incidence (TM polarization). Reprinted with permission from Ref. [23] Copyright 2013 Springer-Nature.

In [29] it was emphasized that the Faraday rotation in the periodic systems is strongly related with the group velocity and gets its maximum values when $v_g$ is zero. In the case of an MPC the dependence of the Faraday angle on the group velocity can be written as follows

$$\Phi_{sp} = \langle Q \rangle \frac{\omega}{2v_g} \qquad (7)$$

where $\langle Q \rangle$ is the matrix element of the MO parameter $\langle Q \rangle = g/\varepsilon$ calculated in the volume of the single lattice cell of the system. Eq. (7) demonstrates the strong correlation between the Faraday effect enhancement and the slow light effect. In the case of plasmonic crystals the mechanism is similar. At the normal incidence the eigenmodes are excited at the Γ point of the Brillouin zone that corresponds to the bandgap edges. The excited modes experience decrease of the group velocity and the effective time of the interaction of a mode with the magnetic media and the conversion to the opposite mode increases, and therefore, the Faraday effect is enhanced.

The experimental demonstration of the Faraday effect enhancement in MPCs similar to the one considered above was done in [23] [Fig. 12(b)]. The spectra of the Faraday rotation exhibit resonant features. The spectra of the Faraday rotation exhibit resonant features. The sample with 495 nm lattice period reaches a maximum Faraday rotation of 0.80° at λ = 963 nm, which is 8.9 times larger than the −0.09° Faraday rotation of the bare iron-garnet film. As seen from Fig.12(c), the same sample shows also a 36% transmittance at the resonant wavelength.

## IV. Nonlinear magnetophotonics

Strong localization and enhancement of electromagnetic fields represents one of the most prominent feature of plasmonics. Obviously, this enhancement can be exploited to boost up the efficiency of a plethora of nonlinear-optical processes, such as, to name a few, Raman scattering or second harmonic generation (SHG), constituting the core of nonlinear plasmonics [163]. The tunability of magnetoplasmonic resonances with external magnetic fields discussed in the previous sections paves the way to the convenient control of the efficiency of nonlinear-optical effects. There, at the crossroads of non-linear optics, plasmonics and magneto-optics, emerges the field of nonlinear magnetoplasmonics aiming at understanding the fundamentals of nonlinear magneto-optics when SPPs are excited.

It's possible to point out a few central aspects of nonlinear magnetoplasmonics. First, of its interest is the tunability of the nonlinear-optical response, attainable in the vicinity of SPP resonances using magnetic fields. Second, it aims at enhancing (otherwise weak) MO effects by means of SPP-driven light localization at the nanoscale. Notably, in both of these approaches the magnetic part of the story is provided by an external DC magnetic field which is used to control the magnetization of the plasmonic medium. This should not be confused with the situations where magnetic field of light at optical frequencies is coupled to the plasmon resonance of medium which does not have to be ferromagnetic. Nonlinear-optical effects associated with the excitation of these magnetic dipole-induced resonances in plasmonic nanostructures [164 -170] will not be discussed here.

Instead, we will overview the possiblities of nonlinear magnetoplasmonics exemplified by magneto-induced SHG (mSHG) as the lowest-order nonlinear-optical process. Most of the formalism shown here is directly applicable to the difference and sum frequency generation (DFG and SFG, respectively) too, which is important, for instance, in THz spectroscopy [171]. The SHG radiation is produced by the nonlinear polarization $P$ at the double frequency $2\omega$ which originates from the anharmonicity of the optical response of the system to the externally applied electromagnetic field $E(\omega)$:

$$P_i(2\omega) = \chi^{(2)}_{ijk}(-2\omega;\omega,\omega)E_j(\omega)E_k(\omega) \qquad (8)$$

where $\chi^{(2)}$ is the second-order nonlinear susceptibility tensor. In magnetized media, the mSHG intensity variations can be characterized by the so-called magnetic contrast $\rho_{2\omega}$:

$$\rho_{2\omega} = \frac{I_{2\omega}(+M)-I_{2\omega}(-M)}{I_{2\omega}(+M)+I_{2\omega}(-M)} \qquad (9)$$

where $I_{2\omega}(\pm M)$ are the SHG intensities measured at the two opposite directions of magnetization $M$. Here, the mSHG intensity variations are governed by the interference of the $P(2\omega)$ contributions originating in the non-magnetic (crystallographic) and magnetization-induced second-order susceptibility tensors, respectively: $\chi^{(2)} = \chi^{(2,cr)} \pm \chi^{(2,m)}$ [172]. Oftentimes, Eq. (9) can be further simplified by considering the ratio of these two (complex) effective susceptibilities $\xi = |\chi^{(2,m)}/\chi^{(2,cr)}|$ and their phase difference $\Delta\varphi$:

$$\rho_{2\omega} = \frac{2|\chi^{(2,cr)}||\chi^{(2,m)}|}{|\chi^{(2,cr)}|^2+|\chi^{(2,m)}|^2}cos\Delta\varphi = \frac{2\xi}{1+\xi^2}cos\Delta\varphi \qquad (10)$$

It is thus clear that the role of SPP resonances on the variations of magnetic contrast can be restricted to their modification of either $\xi$ or $\Delta\varphi$. Indeed, despite boosting the total SHG output, the prominent SPP-induced enhancement of the local fields alone is unable to change $\rho_{2\omega}$, as both crystallographic and magnetic SHG contributions are equally enhanced. Reported rather long ago [173], the first experimental evidence for this might have resulted in delaying the development of nonlinear magnetoplasmonics..

The plasmon-induced variations of $\xi$ can originate in the anisotropy of the $\chi^{(2)}$ tensor. Indeed, the excitation of LSPRs in anisotropic nanostructures results in unequal resonant enhancement of various $\chi^{(2)}$ components responsible for the crystallographic and m SHG, respectively. Absent in spherical nanoparticles, this effect has been demonstrated in anisotropic Ni nanopillars [174]. In the chosen combination of polarizations, crystallographic and magnetic SHG contributions are given by $\chi^{(2)}_{zyy}$ and $\chi^{(2)}_{xyy}$ components, respectively. LSPR modes in these structures facilitates strong enhancement of $E_z(2\omega)$ along the main axis of the pillars, which is equivalent to the resonance in $\chi^{(2)}_{zjk}$ (but not in $\chi^{(2)}_{xjk}$) components. As such, the effective ratio $\xi$ is modified, giving rise to the LSPR-induced variations of $\rho_{2\omega}$. Although this particular system allows for a very clear demonstration of the anisotropy mechanism, the mSHG-LSPR effects in more complicated geometries can be understood in a similar way [175].

At the same time, the experimentally observed propagating SPP-induced variations of the SHG magnetic contrast have been ascribed to the non-locality of the nonlinear-optical response [176-180]. Yet, large number of interfaces and respective non-zero components of the $\chi^{(2)}$ tensor did not allow for a clear picture of relevant non-linear magnetoplasmonic effects in the studied trilayer films. Shortly after, Kirilyuk et al. demonstrated SPP-induced variations of the SFG magnetic contrast in the near-field spectral region [181], pointing out the high promise of this technique for studying magnetic surface excitations.

Preliminary indications of the resonant variations of $\Delta\varphi$ as the origin of the propagating (either on gratings or using prism coupling) SPP-induced mSHG modification were found by Newman et al. [182,183]. It took, however, about a decade until this has been clearly verified by direct SHG interferometry [184,185] and complex polarization analysis [186,187]. Interestingly, similar phase behavior has been reported upon excitation of a collective plasmonic mode in an array of nanodisks [188]. Apparently, the SPP-induced variations of $\Delta\varphi$ and $\xi$ are not always possible to disentangle, as, for example, both are present in many practical situations. For instance, mSHG yield from an isotropic magnetic interface in P-in, P-out combination of polarizations is governed by crystallographic $\chi^{(2)}_{zzz}, \chi^{(2)}_{zxx}, \chi^{(2)}_{xzx} = \chi^{(2)}_{xxz}$ and magneto-induced $\chi^{(2)}_{xzz}, \chi^{(2)}_{xxx}, \chi^{(2)}_{zzx} = \chi^{(2)}_{zxz}$ complex tensor components, so that the effective $\chi^{(2,cr)}$, $\chi^{(2,m)}$ are given by the interference effects. All of them contribute to the total SHG output, resulting in strong intertwining of the amplitude and phase variations originating in the SPP-induced electric field enhancement. Yet, the sign change of magnetic contrast clearly indicates the importance of the SPP-induced phase shift between the $\chi^{(2)}$ components.

Interestingly, the most characteristic feature of nonlinear plasmonics is its sensitivity to the resonances at frequencies different to the fundamental one ($2\omega$ for SHG) [189,190]. This can be exploited for novel mSHG effects where the SPP at the frequency $2\omega$ results in stronger mSHG contrasts than the fundamental SPP resonance [191,192]. Importantly, the system has to support SPP resonances at both $\omega, 2\omega$, which is not the case for purely Au-based structures and typical 1.55 eV photon energy excitation. Large values of $\rho_{2\omega}$ (up to 33%) can be further optimized by adjusting thickness of the plasmonic Ag layer [191] and the excitation wavelength, thus shifting the SPP resonances at the fundamental and SHG frequencies closer to each other due to the SPP dispersion. The latter opens an interesting perspective on the study of resonances overlapping at multiple frequencies to get stronger magnetic modulation of nonlinear-optical effects.

We emphasize the large magnitude of MO effects in nonlinear optics as compared to their linear counterparts. Indeed, if linear magnetoplasmonics typically deals with 0.5-1% reflectivity modulation, in SHG one can quite easily obtain an order of magnitude enhancement. These large effects are not bound to one particular class of objects but are ubiquitous in ferromagnetic metal-based structures, ferromagnetic-noble metal multilayers as well as hybrid noble metal-magnetic dielectric systems [193]. Yet, for sensing and switching applications not only magnetic modulation but also total efficiency of nonlinear-optical conversion is important. However, the strongest SHG modulation is often observed at the minima of the total SHG yield, originating in the destructive interference of multiple contributions. Designing a novel system with overlapping large nonlinear-optical signals and their strong modulation upon magnetization reversal remains one of the open challenges of nonlinear magnetoplasmonics.

## V. Spin-polarization in semiconductors using plasmons

A recent and intriguing development in the field combining plasmons and magnetism, is the extension of the activities towards material systems, incorporating semiconductors. It has been long suggested [194], that future electronic technologies will be relying not only on the control of the charge of the electrons, but also their spin degree of freedom. Already in the 90's, various routes were explored to induce magnetic order or spin-polarization in semiconductors, utilizing light [195- 197]. In these first studies, no particular weight was placed on the effect plasmon resonances might have on the interaction of light with magnetism in semiconductors, as the majority of investigated systems were thin films [195,195]. Indications of interesting physical effects being present in semiconducting particle systems, emerged in works by J. A. Gupta et al. [197] and by R. Beaulac et al. [198]. In the latter, the photomagnetic effects in the form of photoexcited exchange fields leading to strong Zeeman splittings in the band structure, were reported to persist up to room temperature.

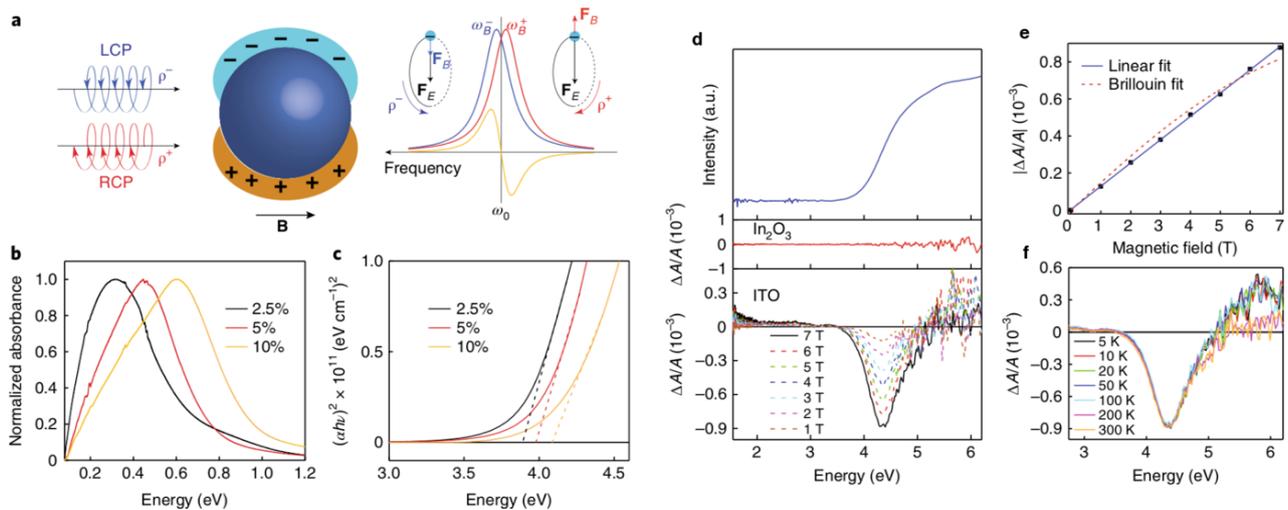

**Fig. 13**. (a) Magnetic circular dichroism (MCD) spectrum related to LSPR in nanoparticles, resulting from the difference in absorption of left (LCP) and right (RCP) circularly polarized light in a magnetic field. (b) LSPR absorption spectra of Indium-Tin Oxide nanocrystals. (c) Tauc plot of the same nanocrystals, used for the determination of the optical bandgaps. (d) MCD spectra expressed as differential absorption (bottom panel), compared to the optical absorption (top panel) and MCD spectrum of $In_2O_3$ nanocrystals. All spectra were measured at 300 K. (e) Field dependence of the MCD intensity, indicating a clear linear dependence. (f) Temperature dependence of the MCD spectra. Adapted with permission from Ref. [202]). Copyright 2018 Springer-Nature.

Existing approaches for achieving spin polarization in semiconductors, have mostly focused on researching dilute magnetic semiconductors or magnetic oxides [199,200]. The approach involving light to achieve spin-polarization or spontaneous magnetization in semiconductors, could potentially add an extra route, adding tunability into the scheme, via the control of size and shape of semiconductor nanoprticles. As an example, ZnO nanocrystals, have been shown to exhibit plasmon resonances in the near-infrared, supported by the observation of a strong magnetic circular dichroism (MCD) signal, which is temperature independent and linearly dependent on the applied magnetic field [201]. More specifically, the temperature independent MCD was attributed to a Pauli-like paramagnetic behavior of the nanocrystals, more common for alkali or noble metals. Recently, non-resonant coupling between cyclotron magnetoplasmonic modes and excitonic states was reported, leading to spin polarization and Zeeman splitting of the excitonic states under externally applied magnetic fields [202] [Fig. 13]. Surprisingly, also for this case the effects persist up to room temperature. Beyond the generation of spin polarized carrier populations in semiconducting materials, recent works have also been reporting on schemes for optical detection of spin currents in hybrid devices. These so far have been based on molecular semiconductors, such as fullerenes [203].

These recent developments open up the way for a fresh look on the plasmon-exciton and plasmon-spin interaction in semiconductors. Combined, they offer the possibility for optical→spin and spin→optical conversion, necessary for a complete framework for an emerging new technology. Expanding this approach also to metal/oxide systems, where plasmon resonances have also been shown to generate and transfer spin currents [204,205], opens up a completely new landscape. Surely, magnetophotonics and magnetoplasmonics will play a crucial role in upcoming developments, concerning material and device designs, holding promises for applicability in information processing and technology [206].

VI. Opto-magnetism: towards an ultrafast control of magnetism and spintronics

The ability to manipulate optical pulses on timescales well into the femtosecond regime has opened the door for attempts to control magnetism in an unprecedented, ultrafast way. In 1996, Beaurepaire and collaborators [207] made a seminal discovery when they observed that a short laser pulse (60 fs, $\lambda$=620 nm) could demagnetize a thin Ni film by 50% on a sub-picosecond timescale. This timescale was much shorter than what was expected from the spin-lattice relaxation at that time. The discovery of the ultrafast quenching of the magnetic order initiated much research and eventually led to the birth of a new scientific branch, *ultrafast magnetism,* poised on the intersection of magnetism and photonics. The microscopic mechanism of the ultrafast suppression of the spin magnetization gave rise to much debate and controversy in the scientific community (see [208,209] for reviews). Obviously, a detailed microscopic understanding of how spin angular momentum can be controlled ultrafast can have far reaching consequences for the future development of ultrafast spintronics, i.e. spintronic devices that can operate at THz frequencies or faster.

While the initial discovery of Beaurepaire *et al.* [207] demonstrated the ultrafast decay of spin moment, a second discovery, made by Stanciu *et al.* [210] showed that optical laser pulses could be used to deterministically reverse the spin moment. Investigating a particular ferri-magnetic alloy, $Gd_{22}Fe_{74.6}Co_{3.4}$, they found that the magnetization direction of magnetic domains could be reverted just by applying continuous radiation or by short laser pulses. This discovery could have important technological implications, since, for example, switching the spin magnetization by an ultrashort pulse could lead to much faster writing of magnetic bits in magnetic recording media. The origin of the all-optical switching (AOS) was initially thought to be linked to the helicity of the laser pulse, i.e., the injected photonic spin moment, but subsequent investigations showed that solely fast laser heating was sufficient to trigger the magnetization reversal [211]. The origin of the helicity-independent switching in GdFeCo alloy was then analyzed to be related to the presence of a magnetization compensation point (antiparallel and nearly compensating moments on Fe, Co and Gd) and the quite different spin-dynamics timescales of the laser excited 3d spin moments on Fe, Co, and the rather slow dynamics of the localized 4f moment on Gd [212,211,213]. Investigating other ferromagnetic compounds and multilayer systems, Mangin and collaborators [214,215] could show that *helicity-dependent all-optical switching* (HD-AOS) was achievable for a broad range of ferromagnetic materials, even for the hard-magnetic recording material FePt that does not exhibit any compensation point. This discovery prompted that there must exist suitable, but as yet poorly known, ways to employ the photon spin angular momentum (SAM) to act on the material's spin moment to trigger spin reversal of the latter.

One of the possible ways for the photon to act on the spin moment could be through an *opto-magnetic effect*, the inverse Faraday effect (IFE). This non-linear MO effect, discovered in the sixties [216], describes the generation of an induced magnetic moment by a circularly polarized electromagnetic wave

$$\boldsymbol{M}^{ind} = \kappa(i\boldsymbol{E} \times \boldsymbol{E}^*) \qquad (11)$$

where $\kappa$ is a materials' dependent constant. The generated magnetization is proportional to the intensity $|E|^2$ and induced along the photon's wavevector. Reversing the helicity from left- to right-circular polarization reverses the direction of the induced magnetization. A first theoretical model for the IFE was proposed by Pitaevskii in the sixties [217]. This model was however based on the assumption that the medium is non-absorbing, a condition which is not met for the metallic materials and nanostructures that have come into the focus in recent years. As it is essential to be able to treat metallic systems, an improved theoretical model that accounts for both effects of photon absorption and photon helicity has been formulated by Battiato *et al.* [218] and Berritta *et al.* [219] (see below). To explain all-optical switching, *dichroic heating* was proposed as an alternative mechanism that could play a role [220]. This mechanism is based on the somewhat different absorption of left and right circularly polarized light in a ferromagnet which implies that the electrons are heated to a somewhat different temperature for left and right circularly polarized radiation, something which could assist switching when the electron temperatures are close to the Curie temperature.

Irrespective of what the deeper origin of the photon-spin interaction is, the spatial resolution of the area where the magnetization could be switched was limited by the light focal spot to domain sizes mostly larger than 10x10 $\mu m^2$. It was consequently realized that, to reach ultrafast light-magnetism operations at the *nanoscale* a further aspect needed to come in. Plasmonics offers precisely the ability to concentrate and enhance electromagnetic radiation much below the diffraction limit, which is essential for opto-magnetic applications in spintronics, where a major goal is deterministic control of nanometer sized magnetic bits. While plasmonics and magnetoplasmonics (see Ref. [5]) have already developed over a longer period, the combination of plasmonics with opto-magnetism represents a new challenge. Circular magnetoplasmonic modes resulting in the shift of the plasmon resonance frequency in Au nanoparticles were investigated by Pineider *et al.* suggesting new detection scheme for label-free refractometric sensing [121]. Thermal effects associated with LSPR in nanoparticles such as hot-electron generation and its dynamics were studied by Saavedra et *al.* [221]. A first attempt to utilize plasmonics to achieve all-optical spin switching on the nanometer scale was made in 2015 by Liu *et al.* [222] who deposited 200 nm Au nano-antennas on a ferrimagnetic TbFeCo film. In this way they could use a high local heating and concentrate the area that exhibits spin switching to about 50 nm diameter. However, it was observed that some areas switched and others didn't. This could be related to a composition inhomogeneity of the TbFeCo film on a sub-100 nm scale. Earlier investigations of AOS in GdFeCo films showed that the switching depends sensitively on the Gd/Fe concentration ratio [223]. An X-ray diffraction study by Graves *et al.* [224] on GdFeCo showed that a composition inhomogeneity could lead to local variations in the switching behavior. A different investigation was made by Kataja *et al.* [225], who could observe both plasmon-induced demagnetization and magnetic switching in a Ni nanoparticle array, excited by a femtosecond laser pulse, which they explained by the plasmonic local heating of the nanoparticles above the Curie temperature. A next step in this direction could be the fabrication of GdFeCo nanoparticle arrays.

To explain the IFE in bulk materials several models have been proposed recently. The deeper understand of the origin of the IFE and how it can be utilized is still the subject of on-going investigations. Hertel developed a plasma model, in which the constant $\kappa$ in Eq. (11) is proportional to $\omega_p^2/\omega^3$ where $\omega_p$ is the plasma frequency and $\omega$ is the frequency of the incident radiation [226]. Nadarajah and Sheldon used this model to estimate the magnetic moment that could be induced in a Au nanoparticle [227]. Popova *et al.* [228] employed a four-level hydrogen model with impulsive Raman scattering to show that such process could lead to a net induced magnetization. Using relativistic electrodynamics Mondal *et al.* showed that there exists an electromagnetic wave-electron spin interaction, which provides a linear coupling of the photon SAM to the electron spin which then acts as an optomagnetic field that generates the induced magnetization [229]. A different approach was developed by Battiato *et al.* [218] and Berritta *et al.* [219], who used second-order density matrix perturbation theory to derive quantum theory expression for the IFE constant $\kappa$ in which the optical absorption is taken into account and that is moreover suitable for *ab initio* calculations. Such calculations are important as they can quantify the size of opto-magnetic interaction. Carrying out *ab initio* calculations for bulk Au, Berritta *et al.* [219] computed that a moment of 7.5 10$^{-3}$ $\mu_B$ per Au atom could be induced by pumping with continuous circular electromagnetic radiation with a 10 GW/cm$^2$ intensity and 800 nm wavelength. Detailed measurements of the light-induced magnetization in Au however do not yet exist. In an early pioneering investigation, Zheludev *et al.* [230] could measure an induced MO polarization rotation of ~7x10$^{-4}$ degree in Au film upon pumping with a laser intensity of 1 GW/cm$^2$ and 1260 nm wavelength, a value that is within an order of magnitude consistent with the *ab initio* calculated induced moment.

The use of plasmonics for the design of strongly enhanced magnetophotonic interactions can now be perceived to be possible in distinct ways. First, plasmonic nanostructures can be tailor-made to concentrate the electromagnetic near field at nano-sized spots, where the enhanced intensity $|E|^2$ can generate a substantial local magnetization $M^{ind}$ via the IFE. It is clear that not only an enhanced intensity is needed, but that the local field must be circularly polarized, too. This implies that a special design of the nanostructures is needed (see, e.g. [231,232]), to ensure that a high magnetic induction pulse is generated. Such modeling can be carried out with Maxwell solvers such as COMSOL or Lumerical. Tsiatmas *et al.* [231] predicted in this way that Ni-Au nanorings excited with a fluence of ~0.1 J/cm$^2$ at plasmon resonance could sustain thermoelectric currents that cause a magnetic induction pulse of ~0.2 T. A different route has been followed by Hurst *et al.* [233], who used a quantum hydrodynamic model to study the magnetic moment induced by circularly polarized radiation in individual Au nanoparticles [Fig. 14(a)]. Circularly polarized radiation can excite electric dipole-like LSPRs in two orthonormal directions on the nanoparticle with a phase difference between them. An orbital magnetization density, $\boldsymbol{m}(t) \propto \boldsymbol{r}(t) \times \frac{d\boldsymbol{r}(t)}{dt}$, appears in the nanoparticle as a result of the free electron motion, leading to a non-vanishing electron current density *J* on the surface of the nanoparticle, see Fig. 14(b). Consequently, the free electron cloud will rotate around the nanoparticle. The magnetic induction *B* due to the circulating current, computed with the Biot-Savart law in the center of the nanoparticle and shown in Fig. 14(c), is predicted to reach 0.3 T for laser intensities of $10^3$ GW/cm$^2$. The magnetic moment *M* induced by this plasmonic IFE can reach ~0.6 $\mu_B$ per Au atom, depending on the size of the nanoparticle and laser intensity, see Fig. 14(d). Even though the assumed laser intensities are very high, the induced moments and generated magnetic fields predicted for the plasmonic IFE [233] of nanoparticles are notably much larger than those computed *ab initio* for the IFE in bulk materials [219]. This strongly increased magnetic moment of the Au nanoparticle nicely illustrates the huge impact that plasmonics could potentially have in the area of opto-magnetism. Specifically, the collective motion of the free electron cloud in the surface plasmon resonance can lead to a much larger total induced magnetization than the excited motions of individual bound electrons.

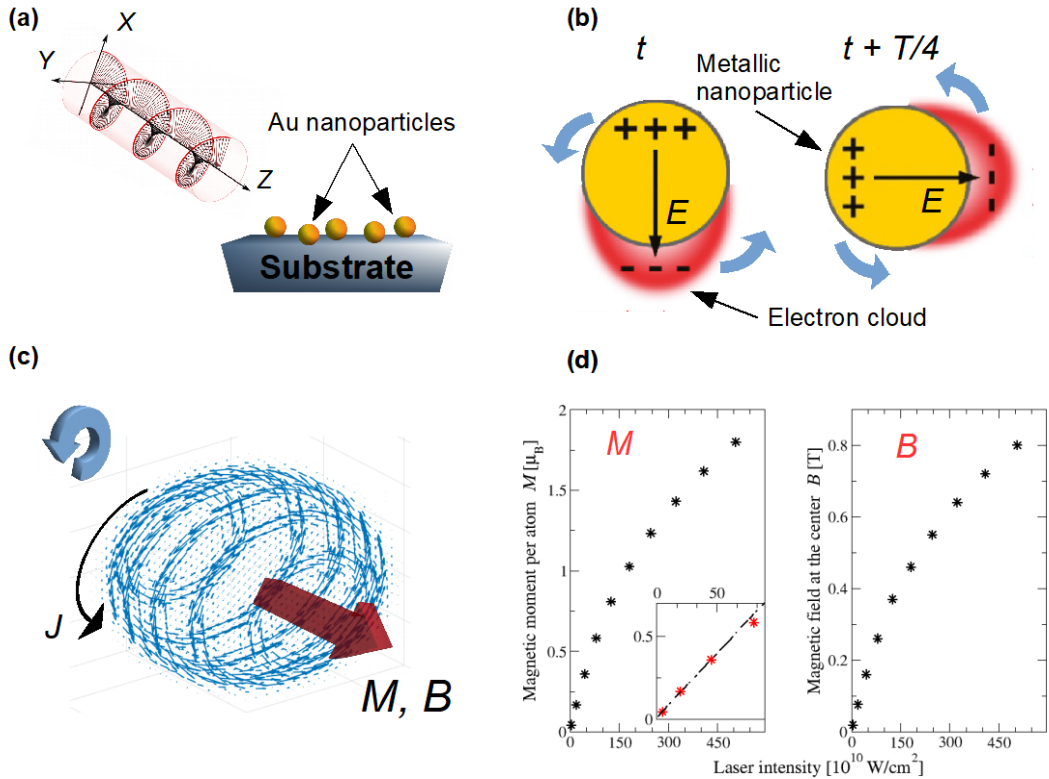

**Fig. 14**. (a) Excitation of gold nanoparticles by circularly polarized coherent radiation with photon energy near the surface plasmon resonance. (b) The rotation of the electromagnetic field vector $E(t)$ induces a rotational motion of the polarized electron cloud with period $T$. (c) A non-vanishing electron current density $J$ on the surface of the nanoparticle arises, which leads to an induced moment $M$ and magnetic induction $B$ that are oriented along the wavevector of the incident radiation. (d) The calculated induced moment per Au atom and the magnetic induction $B$ in the center of the nanoparticle, as a function of the laser intensity. The inset shows that the induced moment increases linearly with the laser intensity, for not too high intensities, evidencing that the induced magnetization is a plasmonic inverse Faraday effect. The calculations were made for a nanoparticle with radius of 1 nm. Reprinted with permission from Ref [233]. Copyright 2018 American Physical Society.

All of the above considerations built upon exploiting the spin angular momentum functionality of the electromagnetic field. A number of years ago it was realized that an optical beam can also carry a well-defined *optical angular momentum* (OAM) [234-237]. For years the OAM has been considered as an exotic, yet benign feature, but more recently its potential usefulness is becoming realized [238, 239]. Beams with high OAM values can nowadays be made in the lab (see e.g. [240-242]). Combining OAM with plasmonics, it has been demonstrated that subfemtosecond dynamics of OAM can be realized in nanoplasmonic vortices [239]. Hence, plasmonic vortices carrying OAM can be confined to deep subwavelength spatial dimension and could offer an excellent time resolution.

The OAM can, therefore, be expected to soon enter the developing area of magnetophotonics, where the OAM could offer a new functionality to control the nanoscale magnetism [135]. There are however many open questions that will have to be solved before this ultimately can be achieved. On short lengthscales comparable to the wavelength of light, the spin angular momentum (SAM) and OAM of a light beam become strongly coupled [243] and it will be difficult to separate their respective contributions. Also, although there is an emerging understanding of the IFE coupling of the SAM of a beam to the electron spin, a similar understanding of the interaction of OAM with spin or orbital magnetism has still to be established. Recently, a first observation of interaction of magnetism and an OAM vortex beam in the THz regime was reported [244]. It can already be perceived that taking both spin and orbital degrees of freedom of photonic beams into account will become paramount for the future development of magnetophotonics.

**Conclusions and future perspectives**

Research on linear and non-linear magnetoplasmonic nanoantennas and nanoscale magnetophotonics has up to now clearly demonstrated the feasibility of active magnetic manipulation of light at the nanoscale. An impact of such active control on applications has been so far hindered by the weak coupling between magnetism and electromagnetic radiation and the high dissipative losses in the used materials. Several strategies, the most promising of which are summarized in this Perspective, have been identified to overcome these limitations. Thereby, this rapidly developing field holds great promise to provide a smart toolbox for actively tunable optical materials and devices in a variety of future disruptive technologies, such as flat nanophotonics, ultrasensitive detection, all-optical and quantum information technologies and spintronics. Nanoscale magnetophotonics could play a prominent role in the design of next-generation technology for computer memory, as the hard disk drive industry is facing a major challenge in continuing to provide increased areal density, driven by the ever-increasing data storage requirements. The heat-assisted magnetic recording approach provides a combination of high coercive field magnetic materials with local heating by a plasmon nanoantenna [245,246]. This approach currently allows up to record-breaking 1 Tb inch$^2$ storage densities. Another practical application of nanoscale magnetophotonics is the use of nanoparticles in medicine, diagnostic techniques and drug delivery due to their potential for direct magnetic manipulation [120,247]. In this regard, solutions of chemically synthetized magnetoplasmonic nanoparticles [248] is fundamental, also in view of potential applications which go beyond nanomedicine, such as the manipulation of the thermal properties of such nanoparticles and/or their environment [249]. Magnetoplasmonic Au-Fe alloy nanoparticles were proved to provide high sensitivity and high resolution in magnetic resonance imaging (MRI), X-ray tomography (CT) and surface enhanced Raman scattering (SERS) [250]. It has been recently demonstrated that magnetochromic hydrogels can be synthetized and be used as magnetic field-modulated color displays [251]. Eventually, multiband MO response would represent another advance in the field [90]. Furthermore, magnetoplasmonic effects can be used for metrology and recently many works pointing in this direction has appeared [89,252,253]. We also foresee that the control of the many degrees of freedom of light (specifically, the optical orbital angular momentum) is within the reach with nanoscale magnetophotonics.

The combination of nanophotonics, magnetoplasmonics and spintronics opens new horizons for practical implementation of magnetic-field controllable nanoscale devices for ultrafast information processing and storage. Newly emerged designs and concepts may help to overcome some of the limitations including plasmon dissipation losses, low efficiency of plasmon excitation in magnetic materials and high magnetic fields required for sufficient modulation. Recent demonstration of tunable multimode lasing modes demonstrated with magnetoplasmonic nanoparticles in combination with organic gain material paves the way for loss-compensated magnetoplasmonic devices [254]. Ultrafast optical excitation also provides means for more efficient excitation of plasmons via the sub-picosecond thermal diffusion of hot electrons due to the formation of nanometer-sized hotspots [255]. Ultrafast control of optical response with spintronics and optical generation of spin waves are very recent advances in the field of nanoscale magnetophotonics, as well. Optical excitation of spin waves [256,257] and optical control of magnetization dynamics [258,259] in GdFeCo and TbFeCo films and magnetic dielectrics by circularly polarized femtosecond laser pulses opens the route for spin wave based devices. Spintronic platforms typically operating with very weak magnetic fields may become next candidates for high-speed photonic devices in mid- and far-IR via the change in resistivity due to the giant magnetoresistance [260]. Local manipulation of the magnetic moments at submicron scale in MO nanodevice with electrically-driven domain wall was recently experimentally implemented [261]. Overall, creating practical magnetophotonics devices will require all-optical and plasmon-assisted control of the magnetic spin and magnetic control of light-matter interactions with low magnetic fields on the nanoscale.

**Acknowledgements**

NM acknowledges financial support from the Luxembourg National Research Fund (FNR CORE Grant No. 13624497 'ULTRON'). IZ acknowledges the financial support from the Knut and Alice Wallenberg Foundation for the postdoctoral grant. IZ and AD acknowledge the Swedish Foundation for Strategic Research (SSF) Future Research Leader Grant. IZ, IAC, VK, PMO and AD acknowledge the Knut and Alice Wallenberg Foundation for the project "Harnessing light and spins through plasmons at the nanoscale" (2015.0060). This work has furthermore been supported by the European Union's Horizon2020 Research and Innovation program, Grant agreement No. 737709 (FEMTOTERABYTE). We thank Michele Dipalo for the help with Fig. 1 and Jerome Hurst for the help with Fig. 14. IR and MPO acknowledge financial support through the CRC/TRR 227. VIB acknowledges the financial support from the Russian Science Foundation (grant No.17-72-20260).